\documentclass[lettersize,journal]{IEEEtran}
\usepackage{algorithmic}
\usepackage{algorithm}
\usepackage{array}
\usepackage[caption=false,font=normalsize,labelfont=sf,textfont=sf]{subfig}
\usepackage{textcomp}
\usepackage{stfloats}
\usepackage{url}
\usepackage{verbatim}
\usepackage{graphicx}
\usepackage{cite}

\usepackage{amsmath,amssymb,amsfonts}
\usepackage{xcolor}
\usepackage{balance}
\def\BibTeX{{\rm B\kern-.05em{\sc i\kern-.025em b}\kern-.08em
    T\kern-.1667em\lower.7ex\hbox{E}\kern-.125emX}}

\begin{document}

\title{Concurrent Entanglement Purification Scheduling in Dynamic Quantum Networks}



\author{Gongyu Ni,~\IEEEmembership{Graduate Student Member,~IEEE},
       Lester Ho,~\IEEEmembership{Senior Member,~IEEE}%
\thanks{This work was supported by Research Ireland under Grant number \text{13/RC/2077\_P2}.}
\thanks{Authors are with Wireless Communications Laboratory, Tyndall National Institute, Dublin D02 W272, Ireland (e-mail: gongyu.ni@tyndall.ie; lester.ho@tyndall.ie).}%
}

\markboth{Journal of \LaTeX\ Class Files,~Vol.~14, No.~8, August~2026}%
{Shell \MakeLowercase{\textit{et al.}}: A Sample Article Using IEEEtran.cls for IEEE Journals}


\maketitle

\begin{abstract}
Quantum networks must deliver end-to-end entanglement whose fidelity meets application requirements, which typically requires entanglement purification along multi-hop paths. Existing scheduling approaches either purify at the link level or optimize end-to-end fidelity for a static set of requests, and do not capture concurrent requests competing for network resources over time. 
This paper studies concurrent entanglement purification scheduling in a time-slotted quantum network with heterogeneous links, in which requests compete for link capacity and quantum memories, and each request may be served by multiple parallel paths for increased reliability. 
We formulate a mixed-integer linear programming (MILP) scheduler that maximizes the number of scheduled requests, with max-min path allocation as a fairness criterion, and develop reinforcement learning (RL) schedulers based on DQN, PPO, and SAC, whose rewards balance request success against Bell-pair consumption. 
Simulations under varying network load, topologies and target fidelity thresholds show that the RL schedulers achieve the highest success rates, improving on MILP by up to 6\% and on Progressive Filling and Greedy heuristics by 16.5\% and 19.5\% respectively, at the cost of higher Bell pair consumption and the deferral of a small fraction of requests. 
MILP, in contrast, achieves the highest fairness and the lowest Bell pair consumption in most scenarios.

\end{abstract}


\section{Introduction}

As an emerging communication technology, quantum networks enable fundamentally new capabilities beyond classical communication. They support secure information exchange through quantum key distribution (QKD)~\cite{bennett2014quantum}, interconnect quantum devices for distributed quantum computing~\cite{cacciapuoti2019}, and improve sensing by coordinating multiple quantum sensors to achieve higher precision~\cite{guo2020distributed}. Beyond these core applications, quantum networks can support a broader range of advanced functionalities, including quantum secure direct communication (QSDC) for transmitting confidential information without prior key establishment~\cite{pan2024evolution}, blind quantum computing for privacy-preserving delegated computation~\cite{fitzsimons2017private}, quantum clock synchronization for high-precision synchronization of remote clocks~\cite{khalid2026quantum}, quantum position verification for authenticating the physical location of network nodes~\cite{quics2026position}, and anonymous quantum transmission for protecting the identities of communicating parties~\cite{liu2024multiparticipant}. All of these applications rely on end-to-end entanglement between remote nodes whose fidelity meets an
application dependent threshold, and many require it to be delivered within bounded time windows.

Establishing such entanglement is challenging. Entangled pairs are degraded by photon loss and channel noise during distribution, decohere while stored in quantum memories, and lose further fidelity at each entanglement swapping operation along a multi-hop path. Entanglement purification \cite{deutsch1996quantum,bennett1996purification,krastanov2019optimized} can restore fidelity by consuming multiple low-fidelity pairs to produce fewer high-fidelity ones, but each round consumes at least two pairs, succeeds only probabilistically, and requires classical communication between the nodes. The number of pairs required therefore grows exponentially with the number of purification rounds, and each round adds latency during which stored qubits continue to decohere. Determining how much purification to perform on each hop thus involves a trade-off between fidelity, resource consumption, and latency.

At the network level, these trade-offs are compounded by contention. Multiple entanglement requests arrive over time and compete for the limited quantum channels and memories of shared links, link availability may vary between time slots, and requests that cannot be accommodated must be deferred. A scheduler must therefore decide which requests to serve, which paths to use, and how many parallel paths to allocate to each request. Because link-level generation and purification are probabilistic, any scheduled path may fail to deliver entangled pairs that meets the fidelity threshold. The scheduling decision therefore determines the balance between request success, Bell-pair efficiency, fairness, and latency.

In this paper, we study concurrent entanglement purification scheduling in a dynamic, time-slotted quantum network. We propose network-level scheduling methods, aiming to allocate entanglement resources for each request such that the resulting fidelity exceeds the required threshold after purification while maximizing the number of requests served within a time slot. The main contributions of this work are summarized as follows:

\begin{itemize}
\item We develop a system model and problem formulation for concurrent entanglement scheduling in a time-slotted network with heterogeneous links. The model captures photon loss, depolarizing and dephasing noise, memory decoherence, imperfect gates and measurements, hop-level purification, and concurrent entanglement swapping. 

\item We develop reinforcement learning (RL)-based scheduling approaches using Deep Q-Network (DQN), Proximal Policy Optimization (PPO), and Soft Actor-Critic (SAC) to maximize entanglement request success rates while accounting for Bell-pair consumption.

\item We formulate a mixed-integer linear programming (MILP) model that maximizes the number of scheduled requests while incorporating residual path allocation as a fairness criterion.

\item We evaluate the proposed methods against Greedy and Progressive Filling (PF) heuristics under varying network loads, topology disruptions, and target fidelity requirements, using latency, request success rate, handling rate, Bell-pair usage efficiency, capacity utilization, and fairness as performance metrics.

\end{itemize}

The remainder of this paper is organized as follows. Section~\ref{sec:related_work} reviews related work. Section~\ref{sec:sys_model} presents the system model, and Section~\ref{sec: prob_formulation} formulates the scheduling problem. Section~\ref{sec: eva_metrics} defines the evaluation metrics, and Section~\ref{sec: sch_methods} describes the scheduling methods. Section~\ref{sec: perf_eval} presents the simulation settings and results, and Section~\ref{sec:conc} concludes the paper.

\section{Related Work}
\label{sec:related_work}
Although various purification protocols have been proposed~\cite{deutsch1996quantum,bennett1996purification,krastanov2019optimized}, existing works primarily focus on link-level purification scheduling~\cite{davies2024entanglement,elsayed2024trade} and lack consideration of end-to-end network-level purification. Research exists on multi-channel entanglement distribution~\cite{vardoyan2024bipartite,chakraborty2020entanglement,zeng2022multi}, but it does not incorporate purification methods to enhance the quality of entanglement establishment. 

In \cite{victora2023entanglement}, the authors integrate entanglement purification into routing across various network topologies, optimizing the hashing rate through path selection and purification. However, their routing heuristics do not jointly account for purification, latency, and dynamic entanglement-request scheduling.

In \cite{li2021effective}, entanglement purification is integrated into routing to support concurrent requests in multi-path, multi-channel quantum networks. The authors develop capacity-allocation strategies, including proportional share, progressive filling, and propagatory update under link-level fidelity constraints. However, the framework assumes discrete processing windows. Queueing across windows, overlapping requests, congestion, and dynamic resource redistribution remain unaddressed. Moreover, purification is performed during network initialization rather than jointly scheduled with individual requests.

The authors of \cite{li2022fidelity} jointly optimizes path selection and purification to maximize the number of entanglement connections satisfying an E2E fidelity requirement. By explicitly modeling E2E fidelity, it identifies critical links for efficient purification. However, it assumes a simplified bit-flip error model and homogeneous Bell-pair fidelity per link. The on-demand Bell-pair generation and time-slotted framework are considered.

To address latency, \cite{chehimi2025entanglement} jointly optimizes average E2E distribution delay and distillation subject to rate and fidelity constraints, incorporating quantum-memory decoherence into the fidelity model. However, its focus is primarily quantum-switch resource allocation rather than joint routing, purification scheduling, and dynamic request handling. It therefore does not determine, which requests to serve, which paths to select, and when or how much purification to perform under time-varying resources.

More recently, \cite{xiao2024purification} proposed a network-level purification scheduling control (PSC) scheme that jointly considers purification and resource allocation to maximize the number of entanglement connections that satisfy the fidelity requirement under E2E fidelity constraints. Nevertheless, it does not fully address purification timing in multi-hop networks, heterogeneous link fidelities, or limited local information.

These limitations motivate a unified routing and scheduling framework that jointly optimizes end-to-end fidelity, latency, and dynamic request handling under time-varying network resources.

\section{System Model}
\label{sec:sys_model}

This section presents a model that integrates quantum protocols and network simulation. Unlike \cite{li2022fidelity}, which follows an advance-generation approach, separating link and network layer functions and focusing routing on resource allocation and path selection, our model adopts an on-demand generation strategy. Entanglement is generated after the routing decision, tightly coupling link and network layers, and jointly considering routing, entanglement distribution, and purification.

\subsection{Quantum error models}

We consider depolarizing noise, dephasing, and photon loss in quantum channels.

\subsubsection{Depolarizing and dephasing errors}

We consider a Bell pair consisting of two qubits. Each qubit experiences independent and identical depolarizing and pure dephasing noise. The single-qubit Lindblad master equation is
\begin{equation}
\frac{d\rho}{dt} = \frac{\gamma_d}{3}\sum_{k=x,y,z}\bigl(\sigma_k\rho\sigma_k - \rho\bigr) \;+\; \frac{\gamma_\phi}{2}\bigl(\sigma_z\rho\sigma_z - \rho\bigr),
\label{eqn:master_equation}
\end{equation}
where \(\rho\) is the single-qubit density matrix, \(\gamma_d \ge 0\) is the depolarizing rate (affects all Pauli errors equally), and \(\gamma_\phi \ge 0\) is the pure dephasing rate (affects only \(\sigma_z\) errors).  
Here \(\sigma_x, \sigma_y, \sigma_z\) are the Pauli errors:
\begin{equation}
\sigma_x = \begin{pmatrix}0&1\\1&0\end{pmatrix},\quad 
\sigma_y = \begin{pmatrix}0&-i\\i&0\end{pmatrix},\quad 
\sigma_z = \begin{pmatrix}1&0\\0&-1\end{pmatrix}.
\end{equation}

Equation~\eqref{eqn:master_equation} generates a Pauli channel on a single qubit. The probabilities for the four Pauli errors \(\{I,X,Y,Z\}\) after time \(t\) are
\begin{equation}
\begin{aligned}
p_I(t) &= \frac{1}{4}\Bigl(1 + e^{-4\gamma_d t/3} + 2e^{-(4\gamma_d/3+\gamma_\phi)t}\Bigr),\\
p_x(t) = p_y(t) &= \frac{1}{4}\Bigl(1 - e^{-4\gamma_d t/3}\Bigr),\\
p_z(t) &= \frac{1}{4}\Bigl(1 + e^{-4\gamma_d t/3} - 2e^{-(4\gamma_d/3+\gamma_\phi)t}\Bigr).
\label{eqn:pauli_error}
\end{aligned}
\end{equation}

Consider a Bell pair whose two qubits evolve independently under the Pauli channel described above. Assuming identical and independent noise processes for both qubits, the corresponding error probabilities are identical, and the fidelity of the resulting two-qubit state is:
\begin{equation}
\label{eqn:channel_fidelity}
\begin{split}
f(t) &= p_I(t)^2 + p_x(t)^2 + p_y(t)^2 + p_z(t)^2 \\
&= \frac{1}{4}\Bigl[1 + e^{-8\gamma_d t/3}
+ 2e^{-(8\gamma_d/3+2\gamma_\phi)t}\Bigr].
\end{split}
\end{equation}
Here, \(\gamma_d\) and \(\gamma_\phi\) denote the depolarizing and dephasing rates per qubit, respectively. 

For transmission through quantum channels, we use Eq.~(\ref{eqn:channel_fidelity}) for a combined depolarizing and dephasing model during transmission. For quantum memories, we assume that storage errors are dominated by pure dephasing and therefore set \(\gamma_d=0\). 

\subsubsection{Photon loss}

To model photon loss during transmission, the loss probability $p_{\text{loss}}$ over a channel is computed using Eq.~(\ref{eq:photon_loss}), where $d$ denotes the physical length of the link between two connected nodes. The fibre attenuation $\alpha$, representing classical optical loss.

\begin{equation}
\label{eq:photon_loss}
p_{\text{loss}} = 1 - 10^{-\frac{d \cdot \alpha}{10}}.
\end{equation}

\subsection{Hop Purification}

For per-hop purification along the node chain between the source and destination nodes, the state remains Bell diagonal under the independent and identical local Pauli channels considered above. Before applying the BBPSSW protocol \cite{bennett1996purification}, we perform a bilateral twirling operation that maps the Bell-diagonal state to a Werner state while preserving its singlet fidelity. The resulting Werner state is written as
\begin{equation}
\label{eqn:entangled_state}
\Phi = p |\phi\rangle\langle\phi| + \frac{1-p}{4} I_4,
\end{equation}
where \(|\phi\rangle\) is the maximally entangled singlet state, \(I_4\) is the two-qubit identity operator, and \(p\in[0,1]\) is the Werner parameter.

Measurement is performed on one of the Bell pairs from every two Bell pairs, and the results are transmitted over a classical channel. The Bell pair is kept only if the measurement results match. If the measurement results differ, it is discarded.


In each hop, under the assumption of the same initial fidelity \(f\) of these generated Bell pairs, the purification fidelity, \(f_p\) , is given by \cite{munro2015inside}:  
\begin{equation}
\label{eqn:hop_fidelity}
f_p= \frac{f^2 + \frac{1}{9} (1-f)^2}{f^2 + \frac{2}{3} f(1-f) + \frac{5}{9} (1-f)^2},
\end{equation}  
with a success probability of purification, \(s_p\) given by:  
\begin{equation}
s_p = f^2 + \frac{2}{3} f(1-f) + \frac{5}{9} (1-f)^2.
\label{eqn:success_probability}
\end{equation}

For multiple purification rounds, $f_1$ is obtained by applying Eq. (\ref{eqn:hop_fidelity}) with $f_0$ as the initial fidelity. Similarly, the purified fidelity $f_2$ is calculated using $f_1$ as the input.

\subsection{Node chain Entanglement}

Entanglement distribution between two end nodes can be achieved either by direct connection over short distances or by quantum repeaters over longer distances. 

\begin{figure}
    \centering
    \includegraphics[width=1\linewidth]{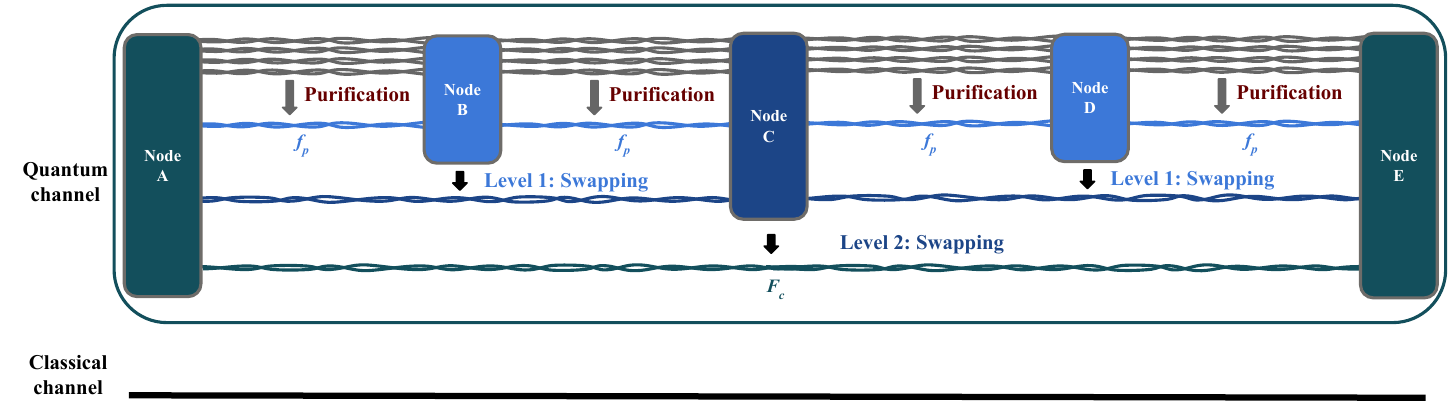}
    \caption{Two-level entanglement distribution. Entangled pairs are first purified within hops $AB$, $BC$, $CD$, and $DE$. First-level entanglement swapping at nodes $B$ and $D$ establishes entanglement across $AC$ and $CE$, respectively. Second-level swapping at node $C$ then establishes end-to-end entanglement between nodes $A$ and $E$.
}
    \label{fig:purified_swapping}
\end{figure}

We assume that entanglement swapping is performed only after purification, following \cite{wang2023efficient}. Each hop undergoes a predetermined number of purification rounds. A node performs swapping once both adjacent hops have completed purification. Hops that finish earlier must wait in quantum memory, accumulating dephasing noise. Swapping follows a concurrent level-swapping scheme (Fig.~\ref{fig:purified_swapping}) : at each level, completed pairs are swapped in parallel, using the swapping time $t_{\mathrm{swap}}$, with any unmatched hop carried forward to the next level until the end nodes are connected.

For a chain of \(n\) neighbouring Werner states across \(n\) connected hops, the output fidelity \(F_c\) is given by \cite{dur1999quantum}:

\begin{equation}
\label{eqn:chain_fidelity}
F_c = \frac{1}{4} + \frac{3}{4} \left[\frac{p_2\left(4 \eta^2 - 1\right)}{3}\right]^{n-1} \prod_{h=1}^{n} \frac{4 f_{p_h} - 1}{3},
\end{equation}
where $p_2$ is the two-qubit gate fidelity, $\eta$ is the measurement fidelity, and $f_{p_h}$ denotes the final fidelity of hop $h$ after purification and all error models.

\subsection{Network model} \label{network_model}

The network model operates in discrete time slots and supports multiple entanglement requests per time slot. Requests that are not selected for service in the current time slot may remain pending for subsequent time slots. Thus, each time slot may contain multiple entanglement requests, including both newly arrived and previously pending requests.

For randomly arriving requests, the number of arrivals in each time slot follows a Poisson distribution, $N_a \sim \mathrm{Poisson}(\lambda)\big|_{[N_{\min},N_{\max}]}$, where $\lambda$ is the network load, $N_{\min}$ and $N_{\max}$ denote the minimum and maximum number of requests per time slot, respectively. The source and destination of the generated requests are selected randomly.

\section{Problem formulation}
\label{sec: prob_formulation}

This paper considers a time-slotted quantum network in which physical links, quantum channels, and quantum-memory resources are available for allocation at the beginning of each scheduling time slot. An entanglement request may be served through one or multiple parallel paths, depending on the available network resources. Each selected path consumes quantum network resources, including quantum-channel capacity, Bell pairs required for entanglement generation and purification, and quantum memories. 

\begin{figure}
\centering
\includegraphics[width=\linewidth]{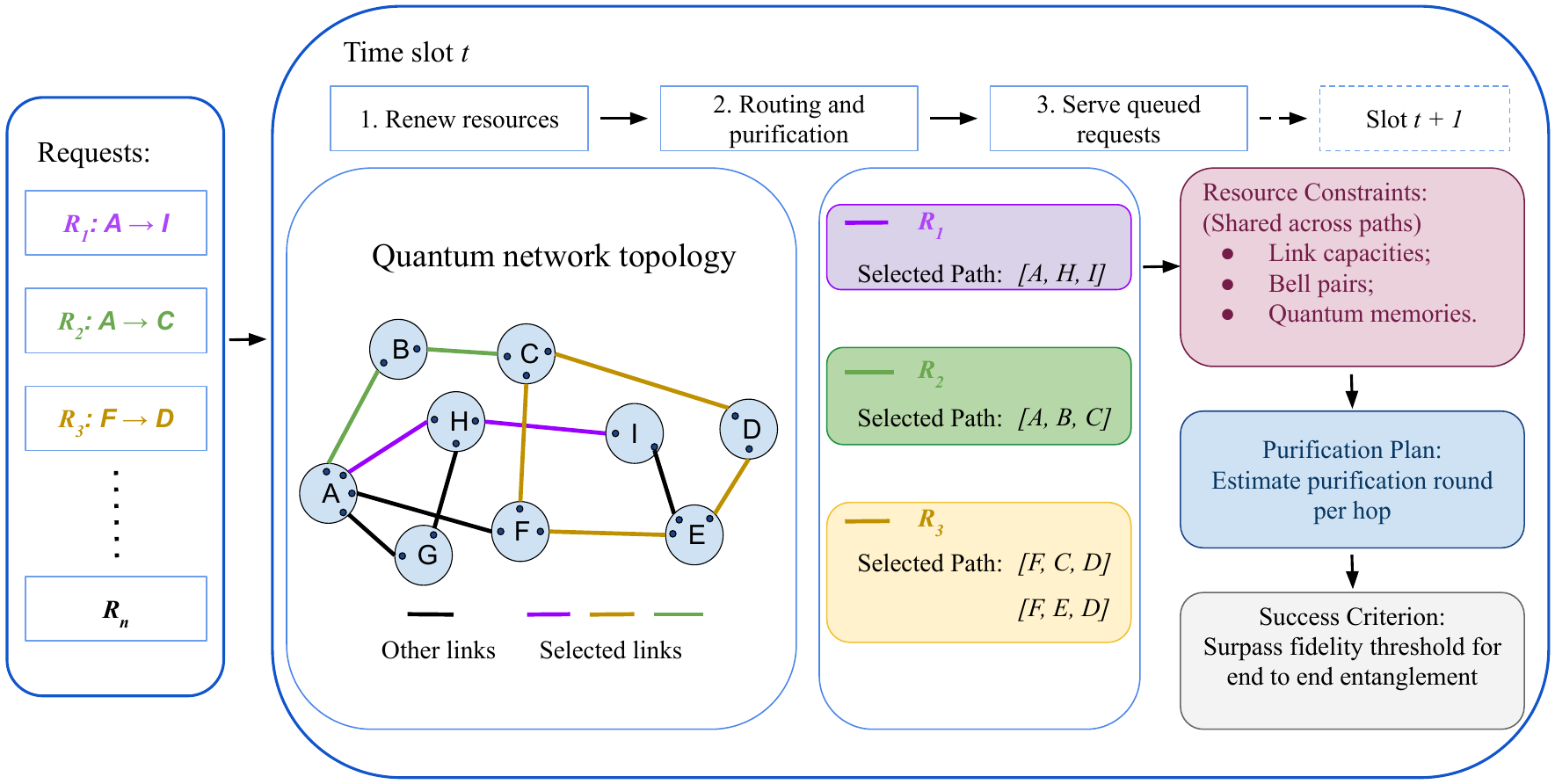}
\caption{Time-slotted quantum network scheduling. Parallel resource-feasible paths can be entangled within one time slot. Multiple paths may serve a single request (e.g., \(R_3\)), providing additional opportunities for successful connection in the presence of propagation and operation errors and purification in the quantum channels.}
\label{fig:purified_entanglement}
\end{figure}

Therefore, the scheduling problem is to select a feasible set of parallel paths for the queued requests while satisfying the required fidelity and the capacity constraints of all shared resources, as shown in Fig. \ref{fig:purified_entanglement}. Requests that cannot be served in the current scheduling slot are deferred to subsequent slots.

\subsection{Quantum Network and Fidelity requirement}

Let the quantum network be modelled as a graph $G=(V,E)$, where $V$ denotes the set of quantum nodes and $E$ denotes the set of physical quantum links. Each link $e\in E$ provides a finite number of quantum channels, while each node $v\in V$ provides a finite number of quantum memories. The available resources are updated at the beginning of each scheduling time slot.

Let \(F\) denote the end-to-end fidelity achieved after purification and entanglement swapping. A request is considered successful if \(F \geq F_\theta\), where \(F_\theta\) is the minimum fidelity threshold, which is dependent on the application.

The scheduling problem therefore determines which candidate paths should be activated for each request such that the selected paths collectively satisfy the network resource constraints while meeting the fidelity requirements.

\subsection{Candidate Paths and Purification Plans} \label{sec:purification}

Let $\mathcal{R}_t$ denote the set of requests considered in time slot $t$, including newly arrived requests and requests deferred from previous scheduling slots. For each request $r\in\mathcal{R}_t$, let
\begin{equation}
\mathcal{P}_r=
\left\{
P_r^1,\ldots,P_r^{K_r}
\right\},
\qquad
K_r\leq K,
\end{equation}
denote the set of candidate paths, where $K$ is the maximum number of candidate paths generated for each request. Multiple candidate paths may be selected simultaneously for the same request.

For each candidate path $P_r^k$, the purification plan determines the number of purification rounds required at each hop. The plan is computed independently for each candidate path according to its required end-to-end fidelity. For a multi-hop path, Eq.~(\ref{eqn:chain_fidelity}) gives the minimum per-hop fidelity assuming all hops achieve the same purified fidelity. The required number of purification rounds at each hop is then determined using the deterministic purification formula in Eq.~(\ref{eqn:hop_fidelity}). Starting from the initial fidelity of each hop, purification is applied iteratively until the hop fidelity reaches the required threshold. The minimum number of rounds satisfying this condition is selected for each hop.


\subsection{Resource Constraints and Path scheduling}

We assume that each physical link \(e\) consists of multiple independent quantum channels. Each channel supports one Bell-pair generation attempt and is associated with one quantum-memory unit at each endpoint, located at the corresponding quantum node, for storing the two qubits of the generated Bell pair.

Let \(C_e(t)\) denote the number of available channels on link \(e\in E\) at time slot \(t\). For candidate path \(P_r^k\), let \(n_{r,k,e}\) denote the number of channels required on link \(e\).

Let $x_{r,k}\in\{0,1\}$ indicate whether candidate path \(P_r^k\) is selected, where \(x_{r,k}=1\) if \(P_r^k\) is scheduled and \(x_{r,k}=0\) otherwise. The link capacity constraint is

\begin{equation}
\sum_{r\in\mathcal R_t}\sum_{k=1}^{K_r}
n_{r,k,e}x_{r,k}
\le C_e(t),
\qquad \forall e\in E.
\label{eq:problem_link_capacity}
\end{equation}

This constraint allows multiple requests, as well as multiple paths belonging to the same request, to share a physical link and node provided that their aggregate resource requirements do not exceed the corresponding capacities.

\section{Evaluation Metrics} 
\label{sec: eva_metrics}

To evaluate different scheduling schemes, we define six metrics: request latency, success rate, handling rate, Bell pair utilization, and capacity utilization. Together, these metrics capture whether requests are served, how quickly and how fairly they are served, and at what cost in network resources.

\subsection{Latency}

The latency $l$ for each successful request is given by  

\begin{equation}  
\label{eqn:request_delay_time}  
    l = t_{f} - t_{s},  
\end{equation}  
where $t_s$ denotes the time at which the request is generated and $t_f$ denotes the time at which an end-to-end Bell pair satisfying the fidelity requirement is delivered. Latency therefore includes both the time spent deferred across scheduling slots and the time required for link-level generation, purification, and entanglement swapping within the slot in which the request is served.
    
\subsection{Success rate}

The success rate $r_s$ is defined as the total number of successfully established entanglement requests $n_s$ divided by total number of requests ${n_{r}}$ in Eq. (\ref{eqn:success_rate}).

\begin{equation}
\label{eqn:success_rate}
r_s = \frac{n_{s}}{n_{r}}.
\end{equation}

The success rate is the primary measure of service quality, as it reflects the fraction of requests that ultimately receive usable entanglement. A request that is scheduled but fails because of photon loss, unsuccessful purification, or decoherence consumes network resources without delivering any benefit to the application. A higher success rate therefore indicates that the scheduler allocates resources to paths that are more likely to meet the fidelity requirement.

\subsection{Handling rate}



The handling rate $r_h$ is the ratio of the number of requests executed in a time slot, $n_e$, to the number of requests considered in that slot, $n_r$, averaged over all time slots:
\begin{equation}
r_h = \frac{1}{N_t} \sum_{t=1}^{N_t} \frac{n_e(t)}{n_r(t)}.
\end{equation}

The handling rate provides an indication of the scheduler's ability to admit concurrent requests within the resource limits of a single slot.
It complements the success rate by distinguishing two causes of unsuccessful service: A lower handling rate indicates that requests are deferred because the scheduler cannot accommodate them, whereas a high handling rate combined with a low success rate indicates that requests are scheduled but fail during execution. A lower handling rate also contributes to higher latency, since deferred requests must wait for subsequent slots.

\subsection{Bell pair utilization rate}

The Bell pair utilization rate $u_b$ is calculated as the ratio of the total number of low-fidelity Bell pairs $n_b$ consumed by the purification process to the total number of successfully established entanglement requests $n_s$ per time slot, averaged over all time slots:
\begin{equation}
\label{eqn:utilization}
u_b = \frac{1}{N_t} \sum_{t=1}^{N_t}\frac{n_{b}(t)}{n_{s}(t)}.
\end{equation}

This metric characterizes how many Bell pairs are used for every successful request. Given the same number of successful requests, a lower Bell pair utilization rate reflects greater efficiency in Bell pair utilization.

\subsection{Capacity Utilization}

The capacity utilization rate $u_c$ is the ratio of the total capacity of the links utilized per slot, $l_u$, to the total available link capacity $L$ in that time slot, averaged over all time slots. 

\begin{equation}
\label{eqn:c_utilization}
u_c = \frac{1}{N_t} \sum_{t=1}^{N_t}\frac{l_{u}(t)}{L(t)}
\end{equation}

A higher capacity utilization indicates that more resources are utilized per slot.
Capacity utilization indicates how much of the network's resources a scheduling scheme occupies, and therefore how much headroom remains for additional traffic. It must be interpreted together with the success rate: high utilization accompanied by a high success rate indicates that resources are used effectively, whereas high utilization with a low success rate indicates that resources are spent on paths that fail. 

\subsection{Scheduling fairness}

The scheduling fairness $J_{\mathrm{req}}$ is defined same as a path fairness in~\cite{li2021effective}, which characterizes the fairness of path allocation among entanglement requests. For each request $r$, the total scheduled path $f_r = \sum_{l \in L_r} f_{r,l}$. The fairness of scheduling across all requests is then calculated as:
\begin{equation}
\label{eqn:fairness}
J_{\mathrm{req}} =
\frac{\left(\sum_{r \in R} f_r\right)^2}
{|R| \sum_{r \in R} f_r^2}.
\end{equation}

The fairness index $J_{\mathrm{req}}$ ranges from $0$ to $1$, where $J_{\mathrm{req}}=1$ indicates that all requests receive an equal amount of scheduled paths, representing complete fairness. 
Fairness matters in multi-user quantum networks, where requests from different users and applications compete for shared links and memories. 
However, fairness is not an objective in itself. Allocating paths evenly can disadvantage requests that need more parallel attempts to succeed, such as those traversing long or low-fidelity routes, so higher fairness may come at the cost of a lower overall success rate. $J_{\mathrm{req}}$ therefore characterizes how each scheme balances equal treatment of requests against the efficient use of shared resources.

\section{Scheduling methods}
\label{sec: sch_methods}

This section presents the scheduling methods evaluated in this paper: a Mixed-Integer Linear Programming (MILP) based scheduler, three Reinforcement Learning (RL) based schedulers using Double DQN, PPO, and SAC, and two heuristic benchmarks, Progressive Filling (PF) and Greedy, adapted from \cite{li2022fidelity} and \cite{li2021effective}, respectively.

All methods operate on the same input in each slot: the residual network $G_t$, the request set $\mathcal{R}_t$, and, for each request, up to $K$ minimal-hop candidate paths with purification plans computed as described in Section~\ref{sec:purification}. Candidate paths whose purification plan cannot reach the required per-hop fidelity within maximum purification rounds are discarded. The methods differ in how they select a subset of the remaining candidates subject to the capacity constraint in Eq.~(\ref{eq:problem_link_capacity}). In all methods, every selected path is executed in the current slot. A request with no selected path is deferred to the next slot.

\subsection{MILP Scheduling}

The MILP scheduler selects candidate paths in order to maximize the number of scheduled requests and, as a secondary objective, the minimum number of paths allocated to any scheduled request. In contrast to \cite{xiao2024purification}, which jointly optimizes purification and resource allocation to maximize the number of connections satisfying the fidelity requirement, the MILP takes the purification plan of each candidate path as given and optimizes the request--path assignment. In particular, it allows multiple parallel paths per request, which increases the request's probability of success.

In addition to the path-selection variables $x_{r,k}$ defined in Section~\ref{sec: prob_formulation}, we introduce a binary variable $z_r$ that indicates whether request \(r\) is completed:
\begin{equation}
z_r
=
\begin{cases}
1, & \text{if request } r \text{ is completed},\\
0, & \text{otherwise}.
\end{cases}
\label{eq:request_completion}
\end{equation}
Note that $z_r = 1$ does not imply that the request is successful, since path outcomes are stochastic and are determined only upon execution. 

Let $\iota_r$ denote the number of candidate paths selected for request \(r\):
\begin{equation}
\iota_r
=
\sum_{k\in\mathcal{P}_r} x_{r,k}
\end{equation}
The request-completion constraints enforce consistency between \(z_r\) and the selected paths:
\begin{equation}
\iota_r\geq z_r,
\label{eq:completion_lower}
\end{equation}
and
\begin{equation}
\iota_r\leq |\mathcal{P}_r|z_r.
\label{eq:completion_upper}
\end{equation}
Therefore, a request is considered completed only when at least one candidate path is selected, while an incomplete request cannot have any selected path.

To incorporate fairness, a continuous variable \(\tau\) is introduced to represent the minimum number of selected paths among completed requests. The fairness constraint is given by
\begin{equation}
\iota_r
\geq
\tau-K(1-z_r),
\qquad
\forall r\in\mathcal{R},
\label{eq:fairness_constraint}
\end{equation}
where \(K\) is the maximum number of candidate paths that can be allocated to a request. 


To ensure that the fairness variable is zero when no request is completed, we additionally impose
\begin{equation}
\tau
\leq
K\sum_{r\in\mathcal{R}} z_r.
\label{eq:fairness_zero}
\end{equation}

The shared link capacities in Eq. (\ref{eq:problem_link_capacity}) are enforced jointly across all selected request-path pairs.


The primary objective is to maximize the number of completed requests, with fairness used as a secondary objective. The resulting MILP can therefore be formulated as:
\begin{subequations}
\begin{align}
\max \quad & W\sum_{r\in\mathcal{R}_t} z_r + \tau \label{eq:milp_main_obj}\\
\text{s.t.} \quad
&\sum_{k\in\mathcal{P}_r}x_{r,k} \ge z_r,&&\forall r \in \mathcal{R}_t,\label{eq:milp_c1}\\
&\sum_{k\in\mathcal{P}_r}x_{r,k} \leq |\mathcal{P}_r|z_r, &&\forall r\in\mathcal{R}_t, \label{eq:milp_c2}\\
& \sum_{k\in\mathcal{P}_r}x_{r,k} \geq \tau-K(1-z_r), &&\forall r\in\mathcal{R}_t, \label{eq:milp_c3}\\
& \tau \leq K\sum_{r\in\mathcal{R}_t}z_r, \label{eq:milp_c4}\\
& \sum_{r\in\mathcal{R}}\sum_{k\in\mathcal{P}_r}n_{r,k,e}x_{r,k}\leq C_e,&&\forall e\in E,\label{eq:milp_c5}\\
& x_{r,k}, z_r \in \{0,1\}, \quad 0 \le \tau \le K. \label{eq:milp_dom}
\end{align}
\end{subequations}
where \(W\) is a sufficiently large positive weight satisfying $W>K$. This choice gives strict priority to increase the number of completed requests and uses the minimum path allocation as a tie-breaking fairness criterion.


After solving the MILP, all selected paths are used to construct the executed-request group. Requests for which no candidate path is selected are placed in the deferred group and can be reconsidered in the next scheduling interval. Algorithm~\ref{alg:milp_scheduler} summarizes the complete MILP scheduling procedure.

\begin{algorithm}
\caption{MILP Scheduling}
\label{alg:milp_scheduler}
\begin{algorithmic}[1]
\STATE \textbf{Input:} Current resource graph \(G_t\), request set \(\mathcal{R}_t\), candidate-path limit \(K\)
\STATE \textbf{Output:} Executed and deferred request groups

\FOR{each request \(r\in\mathcal{R}_t\)}
    \STATE Generate up to \(K\) minimal-hop candidate paths
    \FOR{each candidate path \(P_r^k\)}
        \STATE Determine the purification plan for each hop
        \STATE Compute Bell-pair consumption for each hop
        \STATE Store \(P_r^k\) and its resource-consumption information
    \ENDFOR
\ENDFOR

\STATE Create binary path-selection variables \(x_{r,k}\)
\STATE Create binary request-completion variables \(z_r\)
\STATE Create continuous fairness variable \(\tau\in[0,K]\)
\STATE Add request-completion constraints
\STATE Add minimum-path-allocation fairness constraints
\STATE Add residual link-capacity constraints
\STATE Add the constraint ensuring \(\tau=0\) when no request is completed
\STATE Define the number of completed requests \(T=\sum_{r\in\mathcal{R}_t} z_r\)
\STATE Maximize \(WT+\tau\)

\STATE Solve the MILP optimization problem

\FOR{each request \(r\in\mathcal{R}_t\)}
    \IF{one or more candidate paths are selected}
        \STATE Add all selected paths of \(r\) to the executed group
    \ELSE
        \STATE Add \(r\) to the deferred group
    \ENDIF
\ENDFOR
\STATE \textbf{Return} executed and deferred request groups
\end{algorithmic}
\end{algorithm}

\subsection{Reinforcement learning (RL) based scheduling}

Double DQN (Appendix Section~\ref{sec:apendix_dqn}), PPO (Appendix Section~\ref{sec:apendix_ppo}), and
SAC (Appendix Section~\ref{sec:apendix_sac})-based scheduling employ the same custom
reinforcement learning environment (Appendix Section~\ref{sec:apendix_env}). Their detailed architectures, optimization procedures, and training parameters are provided in the Appendix.

At each scheduling step $i$, the RL agent observes the state $s_i = \{G_i, Z_i, M_i\}$ (Appendix~\ref{sec:apendix_env}), consisting of the residual network graph, the feature matrix of all candidate paths of the requests in $\mathcal{R}_t$, and an action mask identifying the candidates that remain feasible under the residual capacities.
The corresponding RL agent then evaluates the available actions and selects a candidate path according to its learned policy. Thus, although the three algorithms differ in their learning paradigms, they operate on the same state representation, action space, and environmental constraints.

The three approaches are trained using the same environment and epochs to ensure a consistent and fair comparison. They also share the same reward formulation, which combines local allocation performance with global network-level objectives. 

The reward function jointly captures path-level allocation quality and episode-level scheduling performance, enabling all three RL methods to optimize the same scheduling objective. For a selected path $p=P_r^k$, the local reward used to quantify path quality is defined as
\begin{equation}
\label{eq:local_reward}
R_{\mathrm{local}}(p)
=
\operatorname{clip}
\left(
0.05,s(p)
-
0.05\frac{B_{p}}{B_{\mathrm{target}}},
-0.1,0.1
\right),
\end{equation}
where \(s(p)=1\) if the selected path succeeds in establish entanglement surpass fidelity threshold between end nodes, and \(s(p)=-1\) otherwise, \(B_{p}\) denotes the number of Bell pairs consumed by path \(p\) for purification, and \(B_{\mathrm{target}}=160\) is the target Bell-pair budget. A successful path with lower Bell-pair consumption receives a higher reward, reflecting its higher quality and making it more preferable for selection.

To capture the overall scheduling performance of an episode, we define the global reward as
\begin{equation}
\label{eq:global_reward}
R_{\mathrm{global}} =\operatorname{clip} \left(c_{\mathrm{succ}}R_{\mathrm{succ}} + c_{\mathrm{Bell}}R_{\mathrm{Bell}}, -1,1 \right),
\end{equation}
where $R_{\mathrm{succ}}$ is the fraction of requests in $\mathcal{R}_t$
that are successful, and 
\(R_{\mathrm{Bell}}\) is the normalized Bell-pair utilization, given by
\begin{equation}
\label{eq:global_bell}
R_{\mathrm{Bell}}
=\max \left( 1-\frac{B}{B_{\mathrm{target}}}, 0 \right),
\end{equation}
with \(B\) denoting the total Bell-pair consumption in the slot. Since both \(R_{\mathrm{succ}}\) and \(R_{\mathrm{Bell}}\) lie in \([0,1]\), \(c_{\mathrm{succ}}\) and \(c_{\mathrm{Bell}}\) control their relative contributions to the global reward.


In practice, \(c_{\mathrm{succ}}\) should be larger than \(c_{\mathrm{Bell}}\), since unsuccessful requests may still consume Bell pairs. An excessively large \(c_{\mathrm{Bell}}\) can therefore overemphasize resource consumption and weaken the learning signal associated with successful scheduling. Conversely, setting \(c_{\mathrm{succ}}=1\) makes the reward overly focused on request success and provides insufficient incentive to select successful paths with lower Bell-pair consumption. Based on training stability and empirical performance on the training dataset, we set $c_{\mathrm{succ}}=0.8$, and  $c_{\mathrm{Bell}}=0.2$, which provides a stable training process while maintaining a clear trade-off between request success and Bell-pair efficiency.

The RL schedulers and the MILP differ mainly in the information they exploit. The MILP observes only resource capacities and can therefore improve request success only indirectly, by scheduling as many requests as possible and spreading paths evenly among them to promote fairness.  In contrast, the RL agents can learn path quality with the local reward, and is guided by global rewards to directly optimize its behaviour toward maximizing successful requests while controlling Bell-pair utilization, without relying on indirect fairness mechanisms.

The reward associated with transition \(\omega\), which contains the state, action, reward, and next state of each agent, is given by
\begin{equation}
\label{eq:final_reward}
r_\omega = R_{\mathrm{local}}(p_\omega) + \mathbb{I}(\xi_\omega)R_{\mathrm{global}},
\end{equation}
where \(\xi_\omega=1\) indicates that transition \(\omega\) terminates the episode. Consequently, the local reward provides dense path-level feedback at each decision, whereas the global reward is added only once, at the terminal transition, to reflect the overall scheduling outcome.

The relative influence of the two reward components depends on the number of selected paths in an episode. Specifically, if an episode contains \(S_n\) selected paths, the accumulated local contribution is
\begin{equation}
\label{eq:local_episode}
R_{\mathrm{local}}^{\mathrm{ep}}
=
\sum_{\omega=1}^{S_n}
R_{\mathrm{local}}(p_\omega),
\end{equation}
while the global contribution is incurred only once as \(R_{\mathrm{global}}\). Hence, the effective weight of the local objective increases with the number of selected paths. The local reward must therefore remain sufficiently small to avoid diluting the influence of the global scheduling objective when many paths are selected, while still being large enough to provide informative path-level feedback for RL training. Therefore, the local reward scale should be chosen with the typical number of selected paths per episode.

\subsection{Benchmark}

\subsubsection{Progressive-filling}
Progressive filling allocates paths in rounds to equalize the number of paths allocated to each request, following the progressive-filling principle of max-min fairness. For each entanglement request, up to \(K\) candidate paths are generated between its source and destination, together with their corresponding purification plans. The Progressive Filling algorithm \cite{li2021effective} incrementally allocates paths to promote fairness across requests. Initially, each request is assigned zero paths. In each allocation round, requests with fewer than \(K\) allocated paths are considered sequentially. Each request can receive at most \(K\) paths, and the process terminates when no further allocation is feasible.

\subsubsection{Greedy}
The Greedy scheduling algorithm \cite{li2022fidelity} iteratively allocates feasible request--path pairs according to a utility function $U_g$ that balances routing flexibility and resource consumption. For each request, $K$ candidate paths $P_r$ are prepared. The utility of a candidate path is defined as
\begin{equation}
\label{eq:greedy_utility_function}
U_g=\alpha\sum_{v\in P_r}\sum_{u\in\mathcal{N}(v)}\mathbb{I}(c_{vu}) - \beta\sum_{i=1}^{|P_r|-1} C_{\mathrm{pur}}(r_i),
\end{equation}
where $\mathcal{N}(v)$ denotes the set of neighbouring nodes of $v$ along $P_r$, $c_{vu}$ is the residual capacity of link $(v,u)$, and $\mathbb{I}(c_{vu})$ is an indicator that equals $1$ if link $(v,u)$ has positive residual capacity and $0$ otherwise. Here, $r_i$ is the number of purification rounds assigned to hop $i$, and $C_{\mathrm{pur}}(r_i)$ is the corresponding purification resource cost. The parameters $\alpha,\beta\geq0$ control the relative importance of routing flexibility and resource consumption, respectively, and are both set to $0.5$ in this paper.

At each allocation round, the path with the highest utility is iteratively selected. The selected path and its associated purification resources are then removed from the current graph, and the remaining candidate paths are reevaluated based on the updated resources. Each request can be allocated at most $K$ paths. The algorithm terminates when a complete round yields no allocation.

\subsection{Computation complexity analysis}
Let \(N_r\) denote the number of requests, \(K\) the maximum number of candidate paths, \(H\) the maximum number of hops per path, \(M_p\) the maximum number of purification rounds, $N_v$ the number of nodes and \(N_e\) the number of links. Since zero purification rounds are also allowed, each hop has \(M_p+1\) possible purification choices.

\textbf{Benchmark.}
For Progressive Filling and Greedy, the main computational cost arises from evaluating candidate paths. For each candidate path, purification planning and resource evaluation require \(O(H(M_p+1))\) time. Since up to \(K\) candidates are considered for each request across at most \(K\) allocation rounds, the overall candidate-evaluation complexity is \(O\!\left(N_rK^{2}H(M_p+1)\right)\).

\textbf{MILP.}
For each request, up to \(K\) candidate paths are evaluated, giving a preprocessing complexity of $O\left(N_rKH(M_p+1)\right)$, excluding candidate-path generation. The resulting MILP contains \(O(N_rK)\) binary path-selection variables, \(O(N_r)\) binary request-completion variables, one continuous fairness variable, and up to \(O(N_e)\) link-capacity constraints, yielding a model size of $O(N_rK+N_e)$.

Although model construction is polynomial, exact MILP optimization is NP-hard due to the binary path-selection variables and shared link-capacity constraints. With $O(N_rK)$ binary variables, the associated combinatorial search space is exponential, with a worst-case complexity of $2^{O(N_rK)}$. 

\textbf{RL methods.}
For DQN, PPO, and SAC, offline training is excluded from the online scheduling complexity. At each decision step, all three methods process the network graph and \(N_rK\) candidate paths using the same GINE-based graph and path encoders. For a fixed hidden dimension and GINE architecture, the graph encoder costs \(O(N_e+N_v)\), while path encoding and action evaluation cost \(O(N_rKN_v)\). Hence, one inference step has complexity $O\!\left(N_e+N_v+N_rKN_v\right).$ Since paths are selected sequentially, with \(A\le N_rK\) selection steps, the total online inference complexity for DQN, PPO, and SAC is $O\!\left(A\left(N_e+N_v+N_rKN_v\right)\right).$ Thus, in the worst case \(A=O(N_rK)\), the three RL methods have the same asymptotic online scheduling complexity, $O\!\left(N_rK\,N_e+N_r^2K^2N_v\right)$.


\textbf{Summary.}
Table~\ref{tab:complexity} summarizes the scheduling complexities. Overall, Progressive Filling and Greedy have relatively low computational complexity because their main cost is limited to candidate-path evaluation, although they need to repeat this evaluation across multiple allocation rounds. MILP has polynomial model-construction complexity but exponential worst-case optimization complexity due to its \(O(N_rK)\) binary path-selection variables, making it the least scalable approach as the problem size grows. DQN, PPO, and SAC have the same online complexity as they use the same encoders and sequentially evaluate the \(N_rK\) candidate paths. Their online complexity is polynomial and grows with the number of candidate paths, links, and sequential selection steps. Unlike MILP, the three RL methods decision-making uses fixed neural-network inference rather than combinatorial optimization.

\begin{table}[t]
\centering
\caption{Per-slot scheduling complexity}
\label{tab:complexity}
\begin{tabular}{ll}
\hline
Method & Scheduling complexity \\
\hline
PF/Greedy & \(O\!\left(N_rK^{2}H(M_p+1)\right)\) \\
MILP & $O(N_rK+N_e)$ construction; $2^{O(N_rK)}$ worst-case\\
DQN / PPO / SAC & $O\!\left(A\left(N_e+N_v+N_rKN_v\right)\right)$ \\
\hline
\end{tabular}
\end{table}

\section{Performance evaluations}
\label{sec: perf_eval}

In this section, the quantum network simulation settings are presented, and the performance of the scheduling methods are evaluated by metrics defined in Section \ref{sec: eva_metrics}. We first examine the effect of network load, then the effect of topology disruption, and finally the effect of a stricter fidelity requirement.

\subsection{Simulation settings}
\textit{Network topology.} The proposed and benchmark scheduling methods are evaluated on a metropolitan-scale quantum network generated using the Watts--Strogatz model~\cite{watts1998collective}, with $N_v=20$ nodes, mean degree $W_d=6$, and rewiring probability $p_r=0.25$. Each link is independently assigned a distance $d\in\{10,20,30\}$ km, attenuation coefficient $\alpha\in\{0.05,0.15,0.25\}$ dB/km, initial Bell-pair fidelity $F_h^{(0)}\in\{0.80,0.85,0.90,0.95\}$, and capacity $C_h\in\{32,40,48,56,64\}$. The same topology is used for training and evaluation.

\textit{Physical-layer model.} The simulation incorporates imperfect two-qubit gates and measurements, photon loss, transmission depolarization and dephasing, quantum-memory dephasing, entanglement purification, and entanglement swapping, following Section~\ref{sec:sys_model}.

We use \(\bar{\gamma}_{\mathrm{depol}}=5\times10^{-5}\,\mathrm{km}^{-1}\), \(\bar{\gamma}_{\mathrm{deph}}=5\times10^{-6}\,\mathrm{km}^{-1}\), and \(\gamma_m=10^{-5}\,\mu\mathrm{s}^{-1}\) as representative weak-noise benchmark parameters for high-quality fibre links. These values are intended for numerical benchmarking rather than as universal fibre coefficients. Taking the fibre propagation speed as \(v=2\times10^{5}\,\mathrm{km/s}\), the corresponding time-domain rates are \(\gamma_d=10\,\mathrm{s}^{-1}\) and \(\gamma_\phi=1\,\mathrm{s}^{-1}\), with \(\gamma_d=\bar{\gamma}_{\mathrm{depol}}v\) and \(\gamma_\phi =\bar{\gamma}_{\mathrm{deph}}v\). For the memory, \(\gamma_m = 10^{-5} \mu\mathrm{s}^{-1}\) corresponds to a dephasing time \(T_m=1/\gamma_m=100\,\mathrm{ms}\), consistent with experimentally demonstrated photonic-memory coherence times \cite{Koerber2018}.

To model link variability, the initial fidelity $f$ of each hop is perturbed according to
\begin{equation}
\label{eqn:hop_fidelity_distribution}
f\sim\mathcal{N}\left(F_h^{(0)},\sigma_F^2\right),
\qquad
\sigma_F=0.005.
\end{equation}

The initial Bell-pair budget is $N_{\mathrm{bell}}\in\{2,8,16,32\}$ for $r_p\in\{0,1,2,3\}$ purification rounds, respectively. For a hop of distance $d$ and attenuation coefficient $\alpha$, the transmission success probability is
\begin{equation}
\label{eqn:transmission_success}
p_{\mathrm{succ}}=1-p_{\mathrm{loss}},
\end{equation}
where $p_{\mathrm{loss}}$ is given by Eq.~\eqref{eq:photon_loss}. The number of successfully transmitted Bell pairs is sampled as
\begin{equation}
\label{eqn:successful_pairs}
N_{\mathrm{succ}}
\sim
\mathrm{Binomial}\left(N_{\mathrm{bell}},p_{\mathrm{succ}}\right)
\end{equation}
For the node chain entanglement in Eq. \eqref{eqn:chain_fidelity}, the gate and measurement fidelities are set to $p_2=0.998$ and $\eta=0.999$, respectively.

\textit{Timing model.} Each purification attempt succeeds according to Eq.~\eqref{eqn:success_probability}, with a round duration of $t_p=10\mu\mathrm{s}$, assuming classical communication latency below $100\mu\mathrm{s}$ over a 10 km fibre link~\cite{davies2024entanglement}.

After hop-level purification, entanglement swapping is performed recursively. If two input states become available at times $t_1$ and $t_2$, swapping starts at
\begin{equation}
\label{eqn:swap_start_time}
t_{\mathrm{start}}=\max(t_1,t_2),
\end{equation}
and any waiting-induced fidelity degradation follows Eq.~\eqref{eqn:channel_fidelity} with $\gamma_\phi = \gamma_m$ and $\gamma_d=0$. The swapped state becomes available at
\begin{equation}
\label{eqn:swap_ready_time}
t_{\mathrm{ready}}
=t_{\mathrm{start}}+t_{\mathrm{swap}},
\end{equation}
with $t_{\mathrm{swap}}=0.3\mu\mathrm{s}$ \cite{ma2012experimental}. Entanglement generation in quantum sources requires a generation time of $t_g=100\mu\mathrm{s}$, corresponding to a source repetition rate of $R=10$ MHz~\cite{sangouard2011quantum} and assuming $1,000$ attempts to prepare the required number of initial Bell pairs. Hence, the ready time of hop $h$ with $r_p$ purification rounds is
\begin{equation}
\label{eqn:hop_ready_time}
t_h=t_g+r_p t_p.
\end{equation}



\textit{Traffic and dynamics.} Time is divided into $N_t = 1000$ slots of duration $T_{\mathrm{slot}} = 8000~\mu$s. The number of requests arriving in each slot is drawn from a Poisson distribution with mean $\lambda$, truncated to $[N_{\min}, N_{\max}]$, and the source and destination of each request are drawn uniformly at random. To model time-varying resource availability, a fraction $p_d$ of links is removed uniformly at random at the start of each slot and restored at the start of the next. A request is successful if the final end-to-end fidelity satisfies $F \geq F_{\theta}$.

\textit{Scenarios.} The baseline scenario uses $\lambda = 8$ with
$[N_{\min}, N_{\max}] = [5, 12]$, $p_d = 5\%$, and $F_\theta = 0.85$. In each experiment, one parameter is varied relative to the baseline: a lower load of $\lambda = 5$ with $[N_{\min}, N_{\max}] = [3, 8]$, a higher link-removal rate of $p_d = 15\%$, or a stricter fidelity threshold of $F_\theta = 0.90$.

\textit{Implementation.} The RL agents are trained for $10000$ episodes on arrival sequences generated with different arrivals from those used for evaluation. All methods are evaluated on the same sequence of $1000$ slots and request arrivals.The parameters are summarized in Table~\ref{tab:simulation_parameters}.

\begin{table}[htbp]
\addtolength{\tabcolsep}{-9pt}
\caption{Simulation parameters.}
\vspace{-0.1cm}
\renewcommand{\arraystretch}{1.2}
\label{tab:simulation_parameters}
\centering
\begin{tabular}{c c c}
\hline
\hline
\textbf{Parameter} & \textbf{Description} & \textbf{Value} \\
\hline
$N_v$ & Number of network nodes & $20$ \\
$W_d$ & Watts--Strogatz degree & $6$ \\
$p_r$ & Rewiring probability & $0.25$ \\
$d$ & Hop-distance set & ${10,20,30}$ km \\
$\alpha$ & Attenuation set & ${0.05,0.15,0.25}$ dB/km \\
$F_h^{(0)}$ & Initial-fidelity set & ${0.80,0.85,0.90,0.95}$ \\
$C_h$ & Link-capacity set & ${32, 40, 48, 56, 64}$ \\
$p_2$ & Two-qubit gate fidelity & $0.998$ \\
$\eta$ & Measurement fidelity & $0.999$ \\
$\bar{\gamma}_{\mathrm{depol}}$ & Depolarizing rate & $5\times10^{-5}$ km$^{-1}$ \\
$\bar{\gamma}_{\mathrm{deph}}$ & Transmission dephasing rate & $5\times10^{-6}$ km$^{-1}$ \\
$\gamma_{m}$ & Memory dephasing rate & $10^{-5} \mu\mathrm{s}^{-1}$ \\
$v$ & Fibre propagation speed & $2\times10^5$ km/s \\
$t_{\mathrm{swap}}$ & Swapping time & $0.3$ $\mu$s \\
$t_p$ & Purification time per km & $10$ $\mu$s/km \\
$M_p$ & Maximum purification round per hop & 3 \\
$t_g$ & Entanglement-generation time & $100$ $\mu$s \\
$F_{\theta}$ & Fidelity threshold & $0.85$ or $0.90$ \\
$\lambda$ & Network load & $8$ or $5$ requests/slot \\
$N_{\max}$ & Maximum requests per slot & $12$ or $8$\\
$N_{\min}$ & Minimal requests per slot & $5$ or $3$\\
$p_d$ & Dynamic link removal rate & $5 \%$ or $15 \%$\\
$N_t$ & Number of time slots & $1000$ \\
$T_{\mathrm{slot}}$ & Time-slot duration & $8\times10^3$ $\mu\mathrm{s}$ \\
\hline
\hline
\end{tabular}
\end{table}

\subsection{Results}

The Greedy, PF, MILP, DQN, PPO, and SAC methods are evaluated in terms of latency, success rate, capacity utilization, handling rate, Jain's fairness index, and Bell-pair consumption per successful request under varying network loads, dynamic link removal rate and target fidelity threshold.

Latency statistics are computed only over successful requests. Since a deferred request waits at least one slot, each deferral adds approximately $T_{\mathrm{slot}} = 8000~\mu$s to its latency. Latencies below $T_{\mathrm{slot}}$ correspond to requests served in their arrival slot, whereas larger latencies indicate deferral.
 
\subsubsection{Effect of network load}

Tables~\ref{tab:lam5_d5} and~\ref{tab:lam8_d5} compare the scheduling methods under two network loads, $\lambda=5$ and $\lambda=8$, with dynamic link removal rate $p_d=5\%$ and target fidelity threshold $F_{\theta}=0.85$. 


The RL methods achieve the highest success rates in both network load settings. At $\lambda=5$, the DQN method achieved the highest success rate of $0.92$, corresponding to gains of $19.48\%$ over Greedy, $16.46\%$ over Progressive Filling, and $5.75\%$ over MILP. At $\lambda=8$, the DQN and PPO achieved the highest success rate at $0.89$, outperforming MILP by $5.95\%$.

\begin{figure}
\centering
\includegraphics[width=\linewidth]{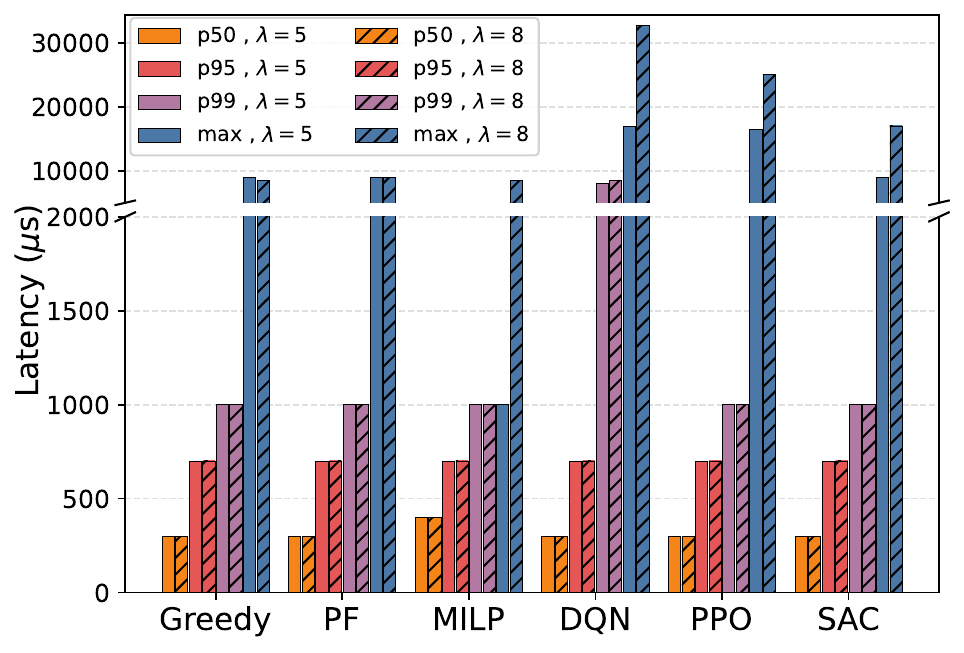}
\caption{Latency percentile comparison for $\lambda=5$ and $\lambda=8$ with $p_d=5\%$ and target fidelity $0.85$.}
\label{fig:combined_delay_percentiles}
\end{figure}

The latency results reveal an important distinction between average and typical request latency. Although the RL methods have relatively higher average latency, their median latency remains close to that of the benchmark methods,as shown in Fig. \ref{fig:combined_delay_percentiles}. At $\lambda=5$, the median latency is $300.3\mu$s for Greedy, PF, DQN, and PPO, and SAC, while MILP has $400.3\mu$s. At $\lambda=8$, the same pattern is observed. 

However, DQN exhibits substantially higher $99^{\text{th}}$ percentile latency, reaching $8100.00\mu$s at $\lambda=5$ and $8500.30\mu$s at $\lambda=8$, compared with approximately $1{,}000.60\mu$s for the other methods, indicating that more than $1\%$ of the requests it serves successfully are deferred by one slot. 
The maximum latencies of DQN and PPO reach $16{,}500$--$17{,}000~\mu$s at $\lambda = 5$, and those of all three RL methods reach $17{,}000$--$32{,}700~\mu$s at $\lambda = 8$, corresponding to deferrals of two to four slots. The higher average latency of the RL
methods is therefore driven by a small number of deferred requests, consistent with their slightly lower handling rates ($0.98$--$0.99$), while the latency of most requests remains comparable to that of the other methods.

MILP and the heuristic benchmarks lack direct knowledge of path quality, so they can improve the success rate only by spreading candidate paths across requests. Accordingly, MILP achieves the highest Jain fairness ($0.99$ at $\lambda = 5$ and $0.98$ at $\lambda = 8$), followed by Greedy ($0.98$, $0.96$) and PF ($0.97$, $0.96$). The RL methods, in contrast, learn which paths are likely to succeed and concentrate additional paths on requests that need multiple parallel attempts, rather than allocating paths evenly to requests that a single path already serves. Their lower fairness ($0.93$ and $0.90$) is therefore the counterpart of their higher success rate.

Higher fairness promotes a more balanced allocation, but can limit the resources available to requests requiring more paths or parallel attempts to succeed.
  
Greedy achieves the lowest average Bell-pair consumption per successful request, requiring $109.72$ Bell pairs at $\lambda=5$, while MILP achieves the lowest consumption of $96.39$ at $\lambda=8$. 
Although a lower Bell-pair utilization rate indicates higher resource efficiency, it is accompanied by a lower success rate. The RL methods consume $9$--$16\%$ more Bell pairs per successful request than MILP, because they allocate additional paths to challenging requests, thereby improving the success rate.

Overall, the RL methods provide consistently higher success rates than Greedy and PF while maintaining similar median latency. MILP provides high fairness, handling rate, and the low Bell-pair consumption per successful request.

\begin{table}[htbp]
\centering
\caption{Performance comparison for $\lambda=5$, $p_d=5\%$, target fidelity $0.85$.}
\addtolength{\tabcolsep}{-3pt}
\label{tab:lam5_d5}
\scriptsize
\begin{tabular}{lcccccc}
\hline
Method & Avg.Lat.($\mu$s) & Success rate & Cap. & Handling rate & Fair & B.P/Success \\
\hline
MILP   & 383.23 & 0.87 & 0.18 & 1.00 & 0.99 & 113.07 \\
DQN    & 448.97 & 0.92 & 0.21 & 0.99 & 0.93 & 127.17 \\
PPO    & 393.28 & 0.91 & 0.20 & 1.00 & 0.93 & 122.92 \\
SAC    & 409.16 & 0.91 & 0.21 & 0.99 & 0.93 & 128.12 \\
PF     & 370.55 & 0.79 & 0.16 & 1.00 & 0.97 & 115.85 \\
GREEDY & 358.74 & 0.77 & 0.15 & 1.00 & 0.98 & 109.72 \\
\hline
\end{tabular}
\end{table}

\begin{table}[htbp]
\centering
\caption{Performance comparison for $\lambda=8$, $p_d=5\%$, target fidelity $0.85$.}
\addtolength{\tabcolsep}{-3pt}
\label{tab:lam8_d5}
\scriptsize
\begin{tabular}{lcccccc}
\hline
Method & Avg.Lat.($\mu$s) & Success rate & Cap. & Handling rate & Fair & B.P/Success \\
\hline
MILP   & 385.35 & 0.84 & 0.24 & 1.00 & 0.98 & 96.39 \\
DQN    & 539.15 & 0.89 & 0.29 & 0.98 & 0.90 & 112.23 \\
PPO    & 437.75 & 0.89 & 0.28 & 0.99 & 0.90 & 106.99 \\
SAC    & 444.61 & 0.88 & 0.29 & 0.99 & 0.90 & 111.92 \\
PF     & 365.69 & 0.79 & 0.25 & 1.00 & 0.96 & 108.53 \\
GREEDY & 352.37 & 0.76 & 0.23 & 1.00 & 0.96 & 105.04 \\
\hline
\end{tabular}
\end{table}



\subsubsection{Varied dynamic link removal rate}

Table~\ref{tab:lam8_d15} compares the methods under a more severe topology perturbation and less network resources, where the dynamic link removal rate $p_d$ increases from $5\%$ to $15\%$ when $\lambda=8$ and target fidelity threshold $F_{\theta}=0.85$. With the same request load but fewer available links in Table~\ref{tab:lam8_d5}, requests compete for a more constrained set of resources. 

Increasing dynamic link removal rate reduces the success rate of all methods. At $p_d=15\%$, the RL methods achieve a success rate of $0.86$, compared with $0.82$ for MILP, $0.76$ for PF, and $0.75$ for Greedy. Relative to Greedy, the RL methods improve the success rate by $14.67\%$. Compared with PF and MILP, the RL methods improve the success rate by $13.16\%$ and $4.88\%$, respectively, demonstrating they still remain higher success under the more disrupted topology.

\begin{figure}
\centering
\includegraphics[width=\linewidth]{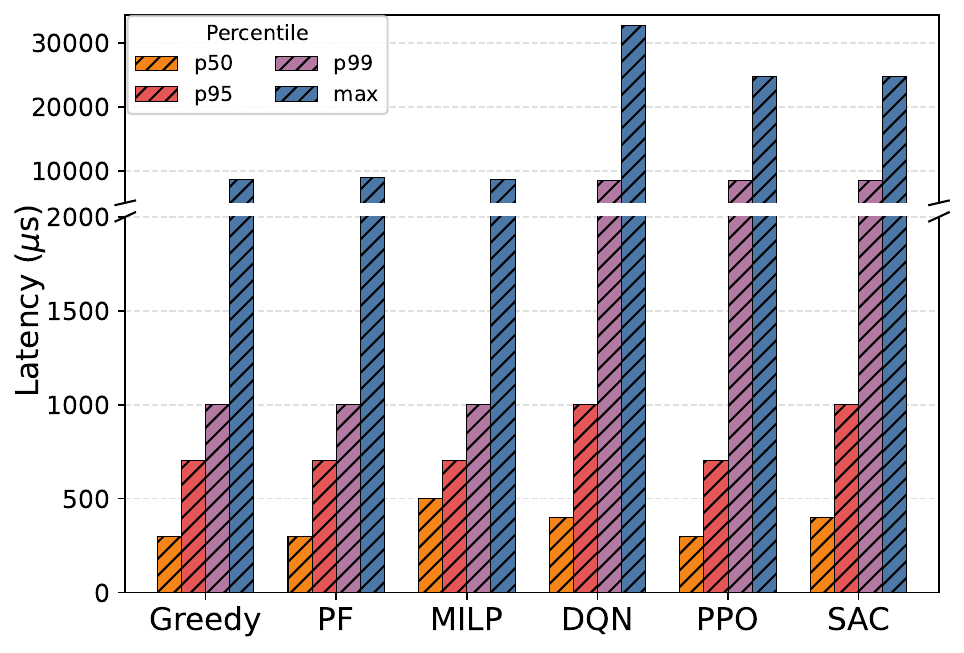}
\caption{Latency percentile comparison for $\lambda=8$, $p_d=15\%$, target fidelity $0.85$.}
\label{fig:rf15_lam8_delay_percentiles}
\end{figure}

Fewer network resources also increase average latency, particularly for the RL methods, reaching $635.36\mu$s for DQN, $525.59\mu$s for PPO, and $570.52\mu$s for SAC, compared with $415.35\mu$s for MILP, $397.44\mu$s for PF, and $377.98\mu$s for Greedy. However, RL methods have median latencies close to those of Greedy and PF ($300.6\mu$s), ranging from $300.6\mu$s to $400.3\mu$s, compared with $500.3\mu$s for MILP. At higher percentiles, DQN, PPO, and SAC exhibit substantially larger delays, with $P99$ reaching approximately $8500\mu$s, compared with about $1000.6\mu$s for the other methods. The maximum latencies are also higher in RL methods, indicating that the larger average latency is mainly driven by a small number of requests experiencing long delays, as shown in Fig. \ref{fig:rf15_lam8_delay_percentiles}.

Capacity utilization increases for all methods, and the RL methods have slightly lower handling rates in $p_d=15\%$ than at $p_d=5\%$, reflecting the increased difficulty of serving requests in the disrupted topology.

MILP achieves the lowest Bell-pair utilization rate at $100.47$ Bell pairs per successful request, followed by PPO ($111.14$). The Bell-pair consumption of the RL methods are similar to Greedy and PF.

Fairness remains higher for the non-RL methods: MILP, Greedy, and PF achieve Jain fairness values of $0.98$, $0.94$, and $0.93$, respectively, compared with around $0.88$ for the RL methods. Overall, the RL methods maintain higher success rates under topology disruption, while MILP provides lower Bell-pair consumption. The benchmark methods achieve similar latency but lower success rates.

\begin{table}[htbp]
\centering
\caption{Performance comparison for $\lambda=8$, $p_d=15\%$, target fidelity $0.85$.}
\addtolength{\tabcolsep}{-3pt}
\label{tab:lam8_d15}
\scriptsize
\begin{tabular}{lcccccc}
\hline
Method & Avg.Lat.($\mu$s) & Success rate & Cap. & Handling rate & Fair & B.P/Success \\
\hline
MILP   & 415.35 & 0.82 & 0.26 & 1.00 & 0.98 & 100.47 \\
DQN    & 635.36 & 0.86 & 0.32 & 0.96 & 0.88 & 114.65 \\
PPO    & 525.59 & 0.86 & 0.31 & 0.98 & 0.89 & 111.14 \\
SAC    & 570.52 & 0.86 & 0.32 & 0.98 & 0.88 & 115.97 \\
PF     & 397.44 & 0.76 & 0.28 & 1.00 & 0.93 & 115.91 \\
GREEDY & 377.98 & 0.75 & 0.26 & 1.00 & 0.94 & 111.90 \\
\hline
\end{tabular}
\end{table}


\subsubsection{Varied target fidelity level}

Table~\ref{tab:rf5_lam8_0.9} compares the methods at a higher target fidelity of $0.90$, while Tables~\ref{tab:lam8_d5} report the corresponding results at target fidelity $0.85$. Increasing target fidelity imposes stricter quality constraints on successful requests, thereby increasing the network resources required to serve them.


Increasing the target fidelity from 0.85 to 0.90 reduces the success rates of MILP, DQN, PPO, and SAC, leaves that of PF unchanged at 0.79, and increases that of Greedy from 0.76 to 0.80. 
The increase in Greedy's success rate is because the higher purification costs change candidate-path rankings through the resource-consumption term in Eq. \eqref{eq:greedy_utility_function}, potentially steering Greedy towards more resource-efficient path combinations. However, this improvement is bounded by the fixed resource budget, as illustrated by the decrease in MILP's success rate from 0.84 to 0.81, bringing its performance closer to that of PF. 

At a target fidelity of 0.90, DQN and PPO achieve the highest success rate of 0.85, representing gains of 6.25\% over Greedy, 7.59\% over PF, and 4.94\% over MILP. Relative to a target fidelity of 0.85, their advantage over PF and Greedy therefore narrows by 5.07 and 10.86 percentage points, respectively, from 12.66\% and 17.11\%. Overall, the stricter fidelity requirement reduces the success rates of MILP and the RL methods, whereas PF is unaffected and Greedy improves.

The higher fidelity requirement also increases capacity utilization for every method. This consistent increase indicates that achieving higher-fidelity requests requires more network resources, leaving fewer resources available for concurrently served requests. Handling rates remain at $1.00$ for MILP, PF, and Greedy, while the RL methods decrease from approximately $0.98$ at fidelity $0.85$ to around $0.93$ at fidelity $0.90$.

\begin{figure}
\centering
\includegraphics[width=\linewidth]{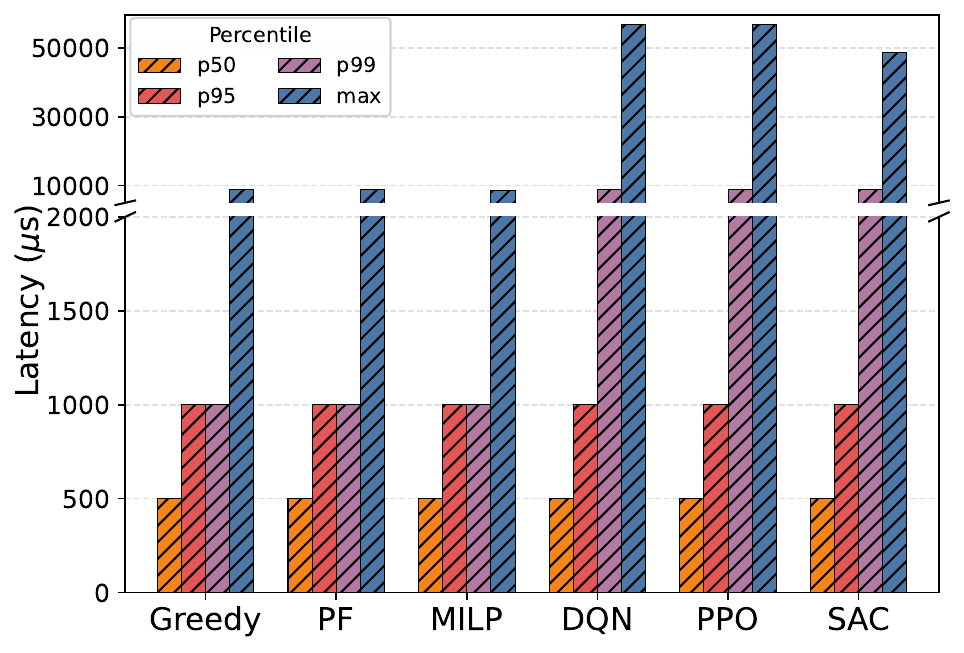}
\caption{Latency percentile comparison for $\lambda=8$, $p_d=5\%$, target fidelity $0.90$.}
\label{fig:rf5_lam8_0.9_delay_percentiles}
\end{figure}

Average latency also increases substantially when the target fidelity is raised. MILP increases from $385.35\mu$s to $503.30\mu$s, while DQN, PPO, and SAC increase from $539.15$, $437.75$, and $444.61\mu$s to $929.56$, $825.40$, and $818.91\mu$s, respectively. PF and Greedy increase from $365.69$ and $352.37\mu$s to $522.38$ and $484.37\mu$s. The latency percentile results in Fig. \ref{fig:rf5_lam8_0.9_delay_percentiles} show a similar trend: the median latency among all the methods are similar. The high-percentile tail also becomes more pronounced for the RL methods, with $P99$ reaching about $9000\mu$s for DQN, PPO, and SAC, compared with approximately $1000\mu$s at fidelity $0.85$. Their maximum latencies increase substantially as well, with DQN and PPO reaching $56700.6\mu$s, corresponding to a deferral of about seven slots.

Bell-pair consumption per successful request increases for every method when the target fidelity is raised from $0.85$ to $0.90$. Fairness decreases at higher fidelity levels across all methods due to fewer resources being available to achieve fairness among requests.

Overall, increasing the target fidelity from $0.85$ to $0.90$ makes request provisioning more resource-intensive: capacity utilization and Bell-pair consumption increase across all methods, average and tail latency increase substantially, and fairness generally decreases. At the same time, success rates decline for most methods, reflecting the additional difficulty of satisfying the stricter fidelity requirement.

\begin{table}[htbp]
\centering
\caption{Performance comparison for $\lambda=8$, $p_d=5\%$, target fidelity $0.90$.}
\addtolength{\tabcolsep}{-3pt}
\label{tab:rf5_lam8_0.9}
\scriptsize
\begin{tabular}{lcccccc}
\hline
Method & Avg.Lat.($\mu$s) & Success rate & Cap. & Handling rate & Fair & B.P/Success \\
\hline
MILP   & 503.30 & 0.81 & 0.29 & 1.00 & 0.96 & 124.30 \\
DQN    & 929.56 & 0.85 & 0.35 & 0.93 & 0.86 & 141.29 \\
PPO    & 825.40 & 0.85 & 0.33 & 0.95 & 0.86 & 134.60 \\
SAC    & 818.91 & 0.84 & 0.34 & 0.95 & 0.86 & 143.55 \\
PF     & 522.38 & 0.79 & 0.31 & 1.00 & 0.87 & 137.35 \\
GREEDY & 484.37 & 0.80 & 0.30 & 1.00 & 0.88 & 132.82 \\
\hline
\end{tabular}
\end{table}


\section{Conclusion}\label{sec:conc}

This paper studied concurrent end-to-end entanglement scheduling with hop-level purification in a time-slotted quantum network, in which multiple requests compete for link capacity and the associated quantum memories, and requests that cannot be served are deferred to subsequent slots. Each request may be served by multiple parallel paths, which act as redundant attempts to deliver a Bell pair that meets the required fidelity threshold. We formulated a MILP scheduler that maximizes the number of scheduled requests, with the minimum path allocation per
request as a secondary fairness objective, and developed RL schedulers based on Double DQN, PPO, and SAC, whose rewards combine path-level success with slot-level request success and Bell-pair consumption. These were compared with Progressive Filling (PF) and Greedy heuristics under varying network load, topology disruption, and fidelity requirements.


The results show that the scheduling objective determines the trade-off between reliability, resource efficiency, fairness, and latency. The RL methods achieved the highest success rate in all scenarios, between $0.84$ and $0.92$, improving on MILP by $4.9$--$6.0\%$ and on the heuristics by $6.3$--$19.5\%$. They achieved this by learning which paths are likely to succeed and concentrating parallel paths on requests that need them, at the cost of $8$--$16\%$ more Bell pairs per successful request than MILP and lower fairness. MILP achieved the lowest Bell-pair consumption per successful request in three of the four scenarios, the highest fairness, and negligible deferral of requests. Greedy achieved the lowest average latency in all scenarios, but also the lowest success rate in
all but the strictest fidelity setting. Although the median latency was similar for all methods, the RL methods deferred a small fraction of requests for one or more slots, raising their 99th-percentile latency to
approximately $8{,}000$--$9{,}000~\mu$s, compared with about $1{,}000~\mu$s for the other methods. A stricter fidelity threshold increased Bell-pair consumption, capacity utilization, and latency for all methods, and narrowed the advantage of the RL methods over the
heuristics.

Several directions follow from these findings. First, incorporating request waiting time or deadlines into the RL state and reward could reduce the deferral tail while retaining the success-rate gains. Second, adding estimated path success probabilities and Bell-pair costs to the MILP objective would provide a stronger optimization baseline and clarify how much of the RL advantage stems from learned path quality. Third, the current path representation depends on the network size, so extending the methods to larger topologies will require size-independent graph representations or multi-agent formulations, together with an evaluation of generalization to unseen topologies. Finally, validation with more detailed hardware models or experimental quantum-network testbeds would establish the practicality of the proposed schedulers under realistic conditions.

\section*{Data Availability}
All data and code required to reproduce all plots shown herein are available at \url{https://doi.org/10.5281/zenodo.23421636}.

\balance
\bibliographystyle{IEEEtran}

\bibliography{ref}

@article{watts1998collective,
  title={Collective dynamics of ‘small-world’networks},
  author={Watts, Duncan J and Strogatz, Steven H},
  journal={nature},
  volume={393},
  number={6684},
  pages={440--442},
  year={1998},
  publisher={Nature Publishing Group}
}

@article{dur1999quantum,
  title={Quantum repeaters based on entanglement purification},
  author={D{\"u}r, Wolfgang and Briegel, H-J and Cirac, Juan Ignacio and Zoller, Peter},
  journal={Physical Review A},
  volume={59},
  number={1},
  pages={169},
  year={1999},
  publisher={APS}
}

@article{bennett1996purification,
  title={Purification of noisy entanglement and faithful teleportation via noisy channels},
  author={Bennett, Charles H and Brassard, Gilles and Popescu, Sandu and Schumacher, Benjamin and Smolin, John A and Wootters, William K},
  journal={Physical review letters},
  volume={76},
  number={5},
  pages={722},
  year={1996},
  publisher={APS}
}

@article{munro2015inside,
  title={Inside quantum repeaters},
  author={Munro, William J and Azuma, Koji and Tamaki, Kiyoshi and Nemoto, Kae},
  journal={IEEE Journal of Selected Topics in Quantum Electronics},
  volume={21},
  number={3},
  pages={78--90},
  year={2015},
  publisher={IEEE}
}

@article{victora2023entanglement,
  title={Entanglement purification on quantum networks},
  author={Victora, Michelle and Tserkis, Spyros and Krastanov, Stefan and de la Cerda, Alexander Sanchez and Willis, Steven and Narang, Prineha},
  journal={Physical Review Research},
  volume={5},
  number={3},
  pages={033171},
  year={2023},
  publisher={APS}
}

@inproceedings{elsayed2024trade,
  title={On the Trade-off between Fidelity and Latency for the Quantum Link Layer with few Memories and Entanglement Purification},
  author={Elsayed, Karim S and KhudaBukhsh, Wasiur R and Rizk, Amr},
  booktitle={2024 International Conference on Quantum Communications, Networking, and Computing (QCNC)},
  pages={17--24},
  year={2024},
  organization={IEEE}
}

@article{davies2024entanglement,
  title={Entanglement buffering with two quantum memories},
  author={Davies, Bethany and I{\~n}esta, {\'A}lvaro G and Wehner, Stephanie},
  journal={Quantum},
  volume={8},
  pages={1458},
  year={2024},
  publisher={Verein zur F{\"o}rderung des Open Access Publizierens in den Quantenwissenschaften}
}

@article{deutsch1996quantum,
  title={Quantum privacy amplification and the security of quantum cryptography over noisy channels},
  author={Deutsch, David and Ekert, Artur and Jozsa, Richard and Macchiavello, Chiara and Popescu, Sandu and Sanpera, Anna},
  journal={Physical review letters},
  volume={77},
  number={13},
  pages={2818},
  year={1996},
  publisher={APS}
}

@article{krastanov2019optimized,
  title={Optimized entanglement purification},
  author={Krastanov, Stefan and Albert, Victor V and Jiang, Liang},
  journal={Quantum},
  volume={3},
  pages={123},
  year={2019},
  publisher={Verein zur F{\"o}rderung des Open Access Publizierens in den Quantenwissenschaften}
}

@article{sangouard2011quantum,
  title={Quantum repeaters based on atomic ensembles and linear optics},
  author={Sangouard, Nicolas and Simon, Christoph and De Riedmatten, Hugues and Gisin, Nicolas},
  journal={Reviews of Modern Physics},
  volume={83},
  number={1},
  pages={33--80},
  year={2011},
  publisher={APS}
}

@article{li2022fidelity,
  title={Fidelity-guaranteed entanglement routing in quantum networks},
  author={Li, Jian and Wang, Mingjun and Xue, Kaiping and Li, Ruidong and Yu, Nenghai and Sun, Qibin and Lu, Jun},
  journal={IEEE Transactions on Communications},
  volume={70},
  number={10},
  pages={6748--6763},
  year={2022},
  publisher={IEEE}
}

@article{cacciapuoti2019,
  title={Quantum internet: networking challenges in distributed quantum computing},
  author={Cacciapuoti, Angela Sara and Caleffi, Marcello and Tafuri, Francesco and Cataliotti, Francesco Saverio and Gherardini, Stefano and Bianchi, Giuseppe},
  journal={IEEE Network},
  volume={34},
  number={1},
  pages={137--143},
  year={2019},
  publisher={IEEE}
}

@article{guo2020distributed,
  title={Distributed quantum sensing in a continuous-variable entangled network},
  author={Guo, Xueshi and Breum, Casper R and Borregaard, Johannes and Izumi, Shuro and Larsen, Mikkel V and Gehring, Tobias and Christandl, Matthias and Neergaard-Nielsen, Jonas S and Andersen, Ulrik L},
  journal={Nature Physics},
  volume={16},
  number={3},
  pages={281--284},
  year={2020},
  publisher={Nature Publishing Group UK London}
}

@article{bennett2014quantum,
  title={Quantum cryptography: Public key distribution and coin tossing},
  author={Bennett, Charles H and Brassard, Gilles},
  journal={Theoretical computer science},
  volume={560},
  pages={7--11},
  year={2014},
  publisher={Elsevier}
}

@article{pan2024evolution,
  title={The Evolution of Quantum Secure Direct Communication: On the Road to the Qinternet},
  author={Pan, Dong and others},
  journal={IEEE Communications Surveys \& Tutorials},
  volume={26},
  number={3},
  pages={1898-1949},
  year={2024},
  publisher={IEEE},
  doi={10.1109/COMST.2024.3367535}
}

@article{fitzsimons2017private,
  title={Private quantum computation: an introduction to blind quantum computing and related protocols},
  author={Fitzsimons, Joseph F.},
  journal={npj Quantum Information},
  volume={3},
  number={1},
  pages={23},
  year={2017},
  publisher={Nature Publishing Group},
  doi={10.1038/s41534-017-0025-3}
}

@article{khalid2026quantum,
  title={Quantum Clock Synchronization Networks: A Survey},
  author={Khalid, Uman and others},
  journal={arXiv preprint},
  year={2026},
  eprint={2604.04437},
  archivePrefix={arXiv},
  primaryClass={quant-ph}
}

@article{quics2026position,
  title={Quantum Position Verification with Remote Untrusted Devices},
  author={QuICS},
  journal={arXiv preprint},
  year={2026},
  eprint={2601.16892},
  archivePrefix={arXiv},
  primaryClass={quant-ph}
}

@article{liu2024multiparticipant,
  title={Multi-participant quantum anonymous communication based on high-dimensional entangled states},
  author={Liu, Jiawei and Mu, Qing-Xia and Che, Ronghua and Wang, Qingle and Han, Yunguang and Cheng, Long},
  journal={Physica Scripta},
  year={2024},
  publisher={IOP Publishing},
  doi={10.1088/1402-4896/ad69d9}
}

@article{xiao2024purification,
  author    = {Xiao, Z. and Li, J. and Xue, K. and et al.},
  title     = {Purification scheduling control for throughput maximization in quantum networks},
  journal   = {Commun. Phys.},
  volume    = {7},
  pages     = {307},
  year      = {2024},
  doi       = {10.1038/s42005-024-01796-2},
  url       = {https://doi.org/10.1038/s42005-024-01796-2}
}

@article{vardoyan2024bipartite,
  title={On the bipartite entanglement capacity of quantum networks},
  author={Vardoyan, Gayane and Van Milligen, Emily and Guha, Saikat and Wehner, Stephanie and Towsley, Don},
  journal={IEEE Transactions on Quantum Engineering},
  volume={5},
  pages={1--14},
  year={2024},
  publisher={IEEE}
}

@article{schulman2017proximal,
  title={Proximal policy optimization algorithms},
  author={Schulman, John and Wolski, Filip and Dhariwal, Prafulla and Radford, Alec and Klimov, Oleg},
  journal={arXiv preprint arXiv:1707.06347},
  year={2017}
}

@article{chakraborty2020entanglement,
  title={Entanglement distribution in a quantum network: A multicommodity flow-based approach},
  author={Chakraborty, Kaushik and Elkouss, David and Rijsman, Bruno and Wehner, Stephanie},
  journal={IEEE Transactions on Quantum Engineering},
  volume={1},
  pages={1--21},
  year={2020},
  publisher={IEEE}
}

@inproceedings{zeng2022multi,
  title={Multi-entanglement routing design over quantum networks},
  author={Zeng, Yiming and Zhang, Jiarui and Liu, Ji and Liu, Zhenhua and Yang, Yuanyuan},
  booktitle={IEEE INFOCOM 2022-IEEE Conference on Computer Communications},
  pages={510--519},
  year={2022},
  organization={IEEE}
}

@article{li2021effective,
  title={Effective routing design for remote entanglement generation on quantum networks},
  author={Li, Changhao and Li, Tianyi and Liu, Yi-Xiang and Cappellaro, Paola},
  journal={npj Quantum Information},
  volume={7},
  number={1},
  pages={10},
  year={2021},
  publisher={Nature Publishing Group UK London}
}

@article{wang2023efficient,
  title={An efficient scheduling scheme of swapping and purification operations for end-to-end entanglement distribution in quantum networks},
  author={Wang, Zhaoying and Li, Jian and Xue, Kaiping and Wei, David SL and Li, Ruidong and Yu, Nenghai and Sun, Qibin and Lu, Jun},
  journal={IEEE Transactions on Network Science and Engineering},
  volume={11},
  number={1},
  pages={380--391},
  year={2023},
  publisher={IEEE}
}

@article{chehimi2025entanglement,
  title={Entanglement distribution delay optimization in quantum networks with distillation},
  author={Chehimi, Mahdi and Goodenough, Kenneth and Saad, Walid and Towsley, Don and Zhou, Tony X},
  journal={IEEE Journal on Selected Areas in Communications},
  volume={43},
  number={5},
  pages={1871--1886},
  year={2025},
  publisher={IEEE}
}

@inproceedings{haarnoja2018soft,
  title={Soft actor-critic: Off-policy maximum entropy deep reinforcement learning with a stochastic actor},
  author={Haarnoja, Tuomas and Zhou, Aurick and Abbeel, Pieter and Levine, Sergey},
  booktitle={International conference on machine learning},
  pages={1861--1870},
  year={2018},
  organization={Pmlr}
}

@inproceedings{van2016deep,
  title={Deep reinforcement learning with double q-learning},
  author={Van Hasselt, Hado and Guez, Arthur and Silver, David},
  booktitle={Proceedings of the AAAI conference on artificial intelligence},
  volume={30},
  number={1},
  year={2016}
}

@article{Koerber2018,
  author  = {K{\"o}rber, Michael and Morin, Olivier and Langenfeld, Sebastian
             and Neuzner, Andreas and Ritter, Stephan and Rempe, Gerhard},
  title   = {Decoherence-protected memory for a single-photon qubit},
  journal = {Nature Photonics},
  volume  = {12},
  pages   = {18--21},
  year    = {2018},
  doi     = {10.1038/s41566-017-0050-y}
}

@article{ma2012experimental,
  title={Experimental delayed-choice entanglement swapping},
  author={Ma, Xiao-song and Zotter, Stefan and Kofler, Johannes and Ursin, Rupert and Jennewein, Thomas and Brukner, {\v{C}}aslav and Zeilinger, Anton},
  journal={Nature Physics},
  volume={8},
  number={6},
  pages={479--484},
  year={2012},
  publisher={Nature Publishing Group UK London}
}

\appendix
\renewcommand{\theequation}{\arabic{equation}}
\setcounter{equation}{0}

\subsection{Custom training Environment} 
\label{sec:apendix_env}

A custom reinforcement learning environment is developed to formulate the sequential entanglement path allocation problem. The environment is responsible for constructing the network state, generating candidate paths, determining purification requirements in the candidate paths and residual link capacities, and updating the network after each path allocation. 

\subsubsection{Input state}

At each decision step, the state consists of three main components: the
current network graph $G_t$, the candidate-path feature matrix $Z_t$, and the action-validity mask $M_t$:
\begin{equation}
\label{eq:dqn_state}
s_t = \left\{G_t,\; Z_t,\; M_t\right\}.
\end{equation}

\paragraph{Graph Representation.}

The quantum network topology is represented by an undirected graph
\begin{equation}
\label{eq:graph}
G_t=(V,E_t),
\end{equation}
where $V$ denotes the set of network nodes and $E_t$ denotes the set of
currently available links.

For each node $v\in V$, the node feature is defined as its graph degree:
\begin{equation}
\label{eq:node_feature}
d_v = \deg_{G_t}(v).
\end{equation}

Each physical link $(u,v)\in E_t$ is associated with three edge features: available link capacity $c_{uv}$, initial fidelity $f_{uv}$, and link distance $d_{uv}$. Therefore, the edge feature vector is
\begin{equation}
\label{eq:edge_feature}
\mathbf{e}_{uv}
=
\left[
c_{uv},\;
f_{uv},\;
d_{uv}
\right].
\end{equation}

\paragraph{Path Feature Representation.}

Each candidate path is represented by a fixed-dimensional feature vector. The node-incidence vector of path $p$ is defined as
\begin{equation}
\label{eq:path_mask}
\mathbf{m}_p
=
[m_1,m_2,\ldots,m_N],
\end{equation}
where
\begin{equation}
\label{eq:path_mask_definition}
m_j=
\begin{cases}
1, & \text{if node }j\text{ belongs to path }p,\\
0, & \text{otherwise}.
\end{cases}
\end{equation}

For each path, purification plan is estimated according to the
required fidelity threshold. $C_{\mathrm{pur}}(p)$ denotes the total purification rounds, while  $R_{\mathrm{pur}}(p)$ is the total Bell pairs consumption of that path.

The maximum number of purification rounds and the fidelity statistics are given by
\begin{equation}
\label{eq:max_rounds}
R_{\max}(p)=\max_i r_i,
\end{equation}
\begin{equation}
\label{eq:min_fidelity}
F_{\min}(p)=\min_i f_i,
\end{equation}
and
\begin{equation}
\label{eq:avg_fidelity}
F_{\mathrm{avg}}(p)
=
\frac{1}{H}
\sum_i f_i,
\end{equation}
where $f_i$ denotes the estimated fidelity associated with the corresponding link of the selected purification plan. $H$ is the total number of hops along the path $p$.

Consequently, the feature vector for candidate path $p$ is defined as
\begin{equation}
\label{eq:path_feature}
\mathbf{z}_p
=
\left[
\mathbf{m}_p,\;
C_{\mathrm{pur}}(p),\;
R_{\mathrm{pur}}(p),\;
R_{\max}(p),\;
F_{\min}(p),\;
F_{\mathrm{avg}}(p)
\right].
\end{equation}


The path-feature matrix is then defined as
\begin{equation}
\label{eq:path_matrix}
Z_t=
\begin{bmatrix}
\mathbf{z}_1\\
\mathbf{z}_2\\
\vdots\\
\mathbf{z}_{N_{\mathrm{path}}}
\end{bmatrix},
\end{equation}
where $N_{\mathrm{path}}$ is the total number of the candidate paths per time slot.

\paragraph{Action Mask.}

A candidate path is considered feasible only when a valid purification plan is available and sufficient capacity exists on every link of the path. The corresponding action mask is
\begin{equation}
\label{eq:action_mask}
M_t=[m_1,m_2,\ldots,m_{N_{\mathrm{path}}}],
\end{equation}
where
\begin{equation}
\label{eq:action_mask_binary}
m_a=
\begin{cases}
1, & \text{if candidate path }a\text{ is currently available},\\
0, & \text{otherwise}.
\end{cases}
\end{equation}

A selected path is removed from subsequent steps, and paths that
become infeasible because of reduced link capacity are also masked out.

\subsubsection{Action}

The Reinforcement learning agent sequentially selects one candidate path at each decision step. The action space is therefore defined as
\begin{equation}
\label{eq:action_space}
\mathcal{A}
=
\left\{
1,2,\ldots,N_{\mathrm{path}}
\right\}.
\end{equation}

Only actions satisfying $m_a=1$ are considered valid. The set of valid
actions at time $t$ is
\begin{equation}
\label{eq:valid_action_set}
\mathcal{A}_t^{\mathrm{valid}}
=
\left\{
a\in\mathcal{A}:m_a=1
\right\}.
\end{equation}

The action selection process terminates when no valid candidate path remains.

\subsection{Deep Q-Network allocation}
\label{sec:apendix_dqn}

Deep Q-Network (DQN) \cite{van2016deep} is employed as a representative value-based reinforcement learning approach for sequential path allocation. The DQN estimates the action value of each candidate path and sequentially selects paths subject to the current feasibility constraints.


\subsubsection{Input state}

At each step $t$, the DQN receives the environment state $s_t$, which consists of the current quantum-network topology and the features of the candidate paths. The graph representation is processed using a graph neural network based on a Graph Isomorphism Network with Edge features (GINE). First, the scalar node feature is projected into a hidden representation of dimension $d=128$. Two GINE convolution layers are then applied to incorporate information from neighbouring nodes and edge attributes. The resulting node representations are aggregated using mean pooling over all nodes to obtain a compact graph-level representation of the current quantum-network state.

In parallel, the candidate path feature matrix is processed independently for each candidate path. For each candidate path, its feature vector is transformed using a fully connected layer to obtain a path-level representation. The graph-level representation and the representations of the candidate paths are then used as the input features for the DQN to evaluate the available actions.






\subsubsection{Action selection}

For each candidate action $a$, the graph-level representation is combined with the corresponding path representation. There are two-layer fully connected Q-value head then estimates the action value. The first layer maps the $256$-dimensional input to $128$ hidden units, and the final layer produces a scalar Q-value for each candidate path.

The DQN agent uses an $\epsilon$-greedy exploration policy. The exploration probability is
\begin{equation}
\label{eq:dqn_epsilon}
\epsilon_t
=
\epsilon_{\min}
+
(\epsilon_{\max}-\epsilon_{\min})
\exp
\left(
-\frac{t}{\tau}
\right),
\end{equation}
with $\epsilon_{\max}=1.0$, $\epsilon_{\min}=0.05$, and $\tau=50000$.

The action selection rule is given by
\begin{equation}
\label{eq:dqn_policy}
a_t=
\begin{cases}
\operatorname{Uniform}
\left(
\mathcal{A}_t^{\mathrm{valid}}
\right),
&
\text{if }u_t<\epsilon_t,
\\[2mm]
\displaystyle
\arg\max_{a\in\mathcal{A}_t^{\mathrm{valid}}}
Q(s_t,a;\theta),
&
\text{otherwise},
\end{cases}
\end{equation}
where $u_t\sim\mathcal{U}(0,1)$.



\subsubsection{Double DQN network}

A target network with parameters $\theta^-$ is maintained separately from the policy network with parameters $\theta$. The target value is defined as
\begin{equation}
\label{eq:double_dqn_target}
\begin{aligned}
y_t &= r_t + \gamma(1-d_t)Q\left(s_{t+1}, a_{t+1}^{*};\theta^{-}\right),\\
a_{t+1}^{*} &= \arg\max_{a} Q(s_{t+1},a;\theta).
\end{aligned}
\end{equation}
where $r_t$ denotes the reward obtained after taking action $a_t$, $\gamma$ is the discount factor, and $d_t\in{0,1}$ indicates whether the transition is terminal ($d_t=1$) or non-terminal ($d_t=0$). The action $a_{t+1}^{*}$ is selected using the policy network with parameters $\theta$, while its value is evaluated using the target network with parameters $\theta^-$. The discount factor is set to $\gamma=0.99$. 

The target network is initialized using the same parameters as the policy network. After each optimization step, the target parameters are updated using a soft update:
\begin{equation}
\label{eq:soft_update}
\theta^-
\leftarrow
\iota\theta + (1-\iota)\theta^-,
\end{equation}
where $\iota=0.01$. This gradual update reduces rapid changes in the target values and improves the stability of Q-learning.

\subsubsection{Loss Function}

The policy network is optimized using the Huber loss:
\begin{equation}
\label{eq:dqn_loss}
\mathcal{L}(\theta)
=
\frac{1}{B}
\sum_{i=1}^{B}
\mathcal{H}_{\delta}
\left(
Q(s_i,a_i;\theta)-y_i
\right),
\end{equation}
where $B=128$ is the mini-batch size, and the Huber loss is defined as
\begin{equation}
\mathcal{H}_{\delta}(x)
=
\begin{cases}
\frac{1}{2}x^2, & |x|\leq\delta,\\[4pt]
\delta\left(|x|-\frac{1}{2}\delta\right), & |x|>\delta,
\end{cases}
\end{equation}
where $\delta>0$ is the Huber loss threshold.

The policy network parameters are updated using the Adam optimizer. To improve optimization stability, the gradient norm is clipped to $\left\|
\nabla_{\theta}\mathcal{L} \right\|_2 \leq 1$.

\subsubsection{Training}

The DQN is trained using transitions generated by sequential path selection. At each environment step, the policy network evaluates all candidate paths, the action mask excludes unavailable paths, and an $\epsilon$-greedy policy selects the next path. The selected path is executed in the environment, after which the network state and path availability are updated. The transition $(s_t,a_t,r_t,s_{t+1},d_t)$ is stored in the replay memory. Once the replay memory contains at least $B=128$ transitions, a random mini-batch is sampled and the policy network is optimized through loss function. The learning rate is further adjusted using a cosine annealing scheduler:
\begin{equation}
\label{eq:cosine_lr}
\eta_t
=
\eta_{\min}
+
\frac{1}{2}
(\eta_0-\eta_{\min})
\left(
1+
\cos
\frac{\pi t}{T_{\max}}
\right),
\end{equation}
where$\eta_0=10^{-5}$, $\eta_{\min}=10^{-7}$, and $T_{\max}=10^4$.

The principal DQN training parameters are summarized in
Table~\ref{tab:dqn_training_configuration}.

\begin{table}
\caption{Parameters used for DQN model training.}
\label{tab:dqn_training_configuration}
\setlength{\tabcolsep}{3pt}
\renewcommand{\arraystretch}{1.2}
\centering
\begin{tabular}{|p{120pt}|p{108pt}|}
\hline
\textbf{Description} & \textbf{Value} \\
\hline
Learning rate &
$1\times10^{-5}$ \\
\hline
Minimum learning rate &
$1\times10^{-7}$ \\
\hline
Batch size &
$128$ \\
\hline
Replay memory size &
$100000$ \\
\hline
Discount factor $\gamma$ &
$0.99$ \\
\hline
Target update coefficient $\iota$ &
$0.01$ \\
\hline
Initial exploration $\epsilon_{\max}$ &
$1.0$ \\
\hline
Minimum exploration $\epsilon_{\min}$ &
$0.05$ \\
\hline
Exploration decay $\tau$ &
$50000$ steps \\
\hline
Gradient clipping &
$1.0$ \\
\hline
\end{tabular}
\end{table}

\subsection{Proximal Policy Optimization allocation}
\label{sec:apendix_ppo}

Proximal Policy Optimization (PPO) \cite{schulman2017proximal} is employed as a representative policy-gradient reinforcement learning approach for path allocation. The objective is to learn a stochastic policy that selects among the currently feasible candidate paths while simultaneously estimating the value of the current network state. The input states, including the graph encoder and path encoder in PPO, are the same as those in Double DQN.

\subsubsection{Actor network}

For each candidate action $a$, the graph representation is concatenated with the corresponding path representation. The actor network then maps this representation to a scalar logit. The first actor layer maps the $256$-dimensional concatenated representation to $128$ hidden units, while the final layer produces one scalar logit for each candidate path.

\subsubsection{Action selection}

The environment provides a binary action-validity mask. The action $a_t$ is sampled from the categorical policy induced by the masked logits, with probabilities defined by the softmax distribution
\begin{equation}
\label{eq:ppo_policy}
\pi_{\theta}(a\mid s_t)
=
\frac{
\exp\left(\widetilde{\ell}_{t,a}\right)
}{
\displaystyle
\sum_{b\in\mathcal{A}_t^{\mathrm{valid}}}
\exp\left(\widetilde{\ell}_{t,b}\right)
}.
\end{equation}



\subsubsection{Critic network}

Unlike the actor, the critic operates only on the graph-level state
representation. The first critic layer maps the $128$-dimensional graph embedding to $128$ hidden units, and the final layer produces a scalar state-value estimate.








\subsubsection{Generalized Advantage Estimation}

The advantage estimates are computed using Generalized Advantage
Estimation (GAE). Let $\widehat{V}_{t}$ denote the critic estimate at time $t$. For each transition, the temporal-difference residual is
\begin{equation}
\label{eq:ppo_delta}
\Delta_t
=
r_t
+
\gamma
(1-d_t)
\widehat{V}_{t+1}
-
\widehat{V}_{t}.
\end{equation}

The advantage estimate is then computed recursively:
\begin{equation}
\label{eq:ppo_gae}
\widehat{A}_t
=
\Delta_t
+
\gamma\lambda
(1-d_t)
\widehat{A}_{t+1},
\end{equation}
where $\gamma=0.99$ and $\lambda=0.95$.

The return target used for critic optimization is
\begin{equation}
\label{eq:ppo_return}
\widehat{R}_t
=
\widehat{A}_t+\widehat{V}_t.
\end{equation}

For the final transition of a non-terminal rollout, the value of the next state is estimated using the current critic. For a terminal final
transition, the next-state value is set to zero.

\subsubsection{PPO clipped objective}

During the PPO update, the probability ratio between the current policy $\pi_{\theta}$ and the policy used to generate the rollout $\pi_{\theta_{\mathrm{old}}}$ is
\begin{equation}
\label{eq:ppo_ratio}
\begin{aligned}
\rho_t(\theta)
&=
\frac{
\pi_{\theta}(a_t\mid s_t)
}{
\pi_{\theta_{\mathrm{old}}}(a_t\mid s_t)
}\\
&=
\exp\left(
\log\pi_{\theta}(a_t\mid s_t)
-
\log\pi_{\theta_{\mathrm{old}}}(a_t\mid s_t)
\right).
\end{aligned}
\end{equation}

The unclipped policy objective is
\begin{equation}
\label{eq:ppo_obj1}
L_t^{\mathrm{CLIP},1}
=
\rho_t(\theta)\widetilde{A}_t,
\end{equation}
while the clipped objective is
\begin{equation}
\label{eq:ppo_obj2}
L_t^{\mathrm{CLIP},2}
=
\operatorname{clip}
\left(
\rho_t(\theta),
1-\epsilon,
1+\epsilon
\right)
\widetilde{A}_t,
\end{equation}
where $\epsilon=0.1$.

The policy loss minimized by the implementation is
\begin{equation}
\label{eq:ppo_policy_loss}
\mathcal{L}_{\mathrm{policy}}
=
\min
\left(
L_t^{\mathrm{CLIP},1},
L_t^{\mathrm{CLIP},2}
\right).
\end{equation}

The clipping operation limits the effective policy update when the new
policy probability deviates too far from the policy that generated the
rollout.

\subsubsection{Value and entropy losses}

The critic is optimized using the mean squared error between the predicted
state value and the GAE return target:
\begin{equation}
\label{eq:ppo_value_loss}
\mathcal{L}_{\mathrm{value}}
=
\operatorname{MSE}
\left(
V_{\phi}(s_t),
\widehat{R}_t
\right).
\end{equation}

The entropy of the categorical policy is
\begin{equation}
\label{eq:ppo_entropy}
\mathcal{H}
\left(
\pi_{\theta}(\cdot\mid s_t)
\right)
=
\sum_{a\in\mathcal{A}_t^{\mathrm{valid}}}
\pi_{\theta}(a\mid s_t)
\log
\pi_{\theta}(a\mid s_t).
\end{equation}

The total loss minimized by the implementation is
\begin{equation}
\label{eq:ppo_total_loss}
\mathcal{L}_{\mathrm{PPO}}
=
\mathcal{L}_{\mathrm{policy}}
+
c_V\mathcal{L}_{\mathrm{value}}
-
c_H\mathcal{H},
\end{equation}
where $c_V=0.1$ and $c_H=0.001$.






\subsubsection{Training}

Training proceeds through repeated request-allocation episodes. At each step, the environment constructs the graph state, candidate-path features, and action mask, while the graph and path encoders generate their respective representations. The actor combines these representations to produce masked action logits, from which a valid path is sampled and executed. The resulting action, old log-probability, value estimate, reward information, and termination flag are stored in the rollout buffer along with the updated state. After an episode ends, its reward is computed and distributed across the episode. Once the buffer contains $1024$ transitions, GAE is computed, advantages are normalized, and the actor and critic are optimized for $3$ epochs with mini-batches of size $64$, using the clipped PPO objective. Finally, the allocation decisions are converted into executed and leftover request groups, with executed requests retaining their selected paths and purification-round information, while unexecuted requests are stored with empty paths.

The principal PPO training parameters are summarized in
Table~\ref{tab:ppo_training_configuration}.

\begin{table}
\caption{Parameters used for PPO model training.}
\label{tab:ppo_training_configuration}
\setlength{\tabcolsep}{3pt}
\renewcommand{\arraystretch}{1.2}
\centering
\begin{tabular}{|p{120pt}|p{108pt}|}
\hline
\textbf{Description} & \textbf{Value} \\
\hline
Learning rate &
$5\times10^{-5}$ \\
\hline
Minimum learning rate &
$1\times10^{-7}$ \\
\hline
Mini-batch size &
$64$ \\
\hline
PPO update threshold &
$1024$ transitions \\
\hline
PPO epochs &
$3$ \\
\hline
Discount factor $\gamma$ &
$0.99$ \\
\hline
GAE parameter $\lambda$ &
$0.95$ \\
\hline
PPO clipping coefficient $\epsilon$ &
$0.1$ \\
\hline
Value-loss coefficient $c_V$ &
$0.1$ \\
\hline
Entropy coefficient $c_H$ &
$0.001$ \\
\hline
Reward decay factor $\delta$ &
$0.99$ \\
\hline
Hidden dimension &
$128$ \\
\hline
Gradient clipping &
$1.0$ \\
\hline
Cosine scheduler $T_{\max}$ &
$200$ \\
\hline
\end{tabular}
\end{table}

\subsection{Soft Actor-Critic allocation}
\label{sec:apendix_sac}

Soft Actor-Critic (SAC) \cite{haarnoja2018soft} is also employed as an reinforcement learning approach for sequential path allocation. SAC consists of one stochastic actor, two independent critic networks, two corresponding target critic networks, and an automatically adjusted entropy coefficient. The input states, including the graph encoder and path encoder in SAC, are the same as those in PPO and Double DQN.

\subsubsection{Actor network}

The actor maps the combined graph-path representation to one scalar logit for each candidate path. Invalid candidate paths are masked before constructing the policy distribution. The log-policy probability used by the SAC objective is
\begin{equation}
\label{eq:sac_log_policy}
\log\pi_\theta(a\mid s_t)
=
\log
\left[
\pi_\theta(a\mid s_t)
\right].
\end{equation}

During environment interaction, the next path is sampled from the categorical policy $a_t \sim \pi_\theta(\cdot\mid s_t)$.

\subsubsection{Twin critic networks}

Two independent critics are used to estimate the action value of every candidate path. Each critic uses the graph-path representation and the same two-layer fully connected head. For the action selected during interaction, the corresponding critic values are $Q_{\phi_1}(s_t,a_t)$, $Q_{\phi_2}(s_t,a_t)$. The minimum of the two critic estimates is used to reduce overestimation:
\begin{equation}
\label{eq:sac_q_min}
Q_{\min}(s_t,a)
=
\min
\left(
Q_{\phi_1}(s_t,a),
Q_{\phi_2}(s_t,a)
\right).
\end{equation}

\subsubsection{Target critic networks}

Two target critics,
$Q_{\bar{\phi}_1}$ and $Q_{\bar{\phi}_2}$, are maintained separately. For the next state $s_{t+1}$, the target critics produce action-value estimates for all candidate paths. Their minimum is defined as

\begin{equation}
\label{eq:sac_target_q_min}
Q_{\min}^{\mathrm{target}}
(s_{t+1},a)
=
\min
\left(
Q_{\bar{\phi}_1}(s_{t+1},a),
Q_{\bar{\phi}_2}(s_{t+1},a)
\right).
\end{equation}


\subsubsection{Entropy-regularized target}

The soft state value used to construct the critic target is the expected entropy-regularized action value:
\begin{equation}
\label{eq:sac_soft_value}
\begin{aligned}
V(s_{t+1})
&=
\sum_a \pi_\theta(a\mid s_{t+1})
\Big[
Q_{\min}^{\mathrm{target}}(s_{t+1},a)
\\
&\qquad\qquad
-\alpha\log\pi_\theta(a\mid s_{t+1})
\Big].
\end{aligned}
\end{equation}

The resulting target for the critic is
\begin{equation}
\label{eq:sac_target}
y_t
=
r_t
+
(1-d_t)
\gamma
V(s_{t+1}),
\end{equation}
where $d_t=1$ denotes a terminal transition and $\gamma=0.99$.

\subsubsection{Critic loss}


The two critics are optimized using the Huber loss.
\begin{equation}
\label{eq:sac_critic_loss1}
\mathcal{L}_{Q_1}
=\mathcal{H}_{\delta}\left(
Q_{\phi_1}(s_t,a_t)-y_t
\right),
\end{equation}
\begin{equation}
\label{eq:sac_critic_loss2}
\mathcal{L}_{Q_2}
=\mathcal{H}_{\delta}\left(
Q_{\phi_2}(s_t,a_t)-y_t
\right).
\end{equation}

\subsubsection{Actor loss}

The actor maximizes the entropy-regularized expected action value. Since the implementation performs gradient descent, the corresponding actor loss is

\begin{equation}
\label{eq:sac_actor_loss}
\mathcal{L}_{\mathrm{actor}}
=
\sum_a
\pi_\theta(a\mid s_t)
\left[
\alpha\log\pi_\theta(a\mid s_t)
-
Q_{\min}(s_t,a)
\right].
\end{equation}

Minimization of actor loss encourages the policy to
increase the probability of actions with high estimated value while maintaining entropy through the temperature coefficient $\alpha$.

\subsubsection{Entropy and temperature loss}

The policy entropy is calculated from the categorical action
distribution:
\begin{equation}
\label{eq:sac_entropy}
\mathcal{H}
(\pi_\theta(\cdot\mid s_t))
=
\sum_a
\pi_\theta(a\mid s_t)
\log\pi_\theta(a\mid s_t).
\end{equation}

The entropy coefficient is parametrized through its logarithm $\alpha = \exp(\log\alpha)$. The initial entropy coefficient is $\alpha_0=0.2$. The target entropy in the implementation is $\mathcal{H}_{\mathrm{target}}=-1.0$.

The temperature loss implemented in the SAC agent is
\begin{equation}
\label{eq:sac_alpha_loss}
\mathcal{L}_{\alpha}
=
\log\alpha
\left(
\mathcal{H}
(\pi_\theta(\cdot\mid s_t))
-
\mathcal{H}_{\mathrm{target}}
\right).
\end{equation}




\subsubsection{Soft target-network update}

After each critic update, the online critic parameters are transferred
gradually to their target networks using a soft update:
\begin{equation}
\label{eq:sac_soft_update}
\bar{\phi}_i
\leftarrow
\tau\phi_i
+
(1-\tau)\bar{\phi}_i,
\qquad
i\in{1,2},
\end{equation}

where $\tau=0.005$. The use of a small value of $\tau$ reduces rapid variation in the target values and improves the stability of the off-policy critic updates.

\subsubsection{Training}

The SAC agent is trained on sequential transitions generated during candidate-path allocation. At each step, the actor evaluates all candidate paths, applies the action-validity mask, and samples a valid path, which is then executed by the environment to obtain the next state and transition information. The episode reward is converted to the decayed reward and stored in the replay buffer. The replay buffer has a maximum capacity of $100000$. Once the replay buffer contains at least $128$ transitions, mini-batches are sampled randomly for optimization. Each mini-batch contains $32$ transitions. For each iteration, the target value is computed using the target critics and entropy-regularized policy in Eq.~\eqref{eq:sac_target}, followed by critic, actor, and entropy-coefficient updates. Finally, the target critics are updated according to Eq.~\eqref{eq:sac_soft_update}.

The principal SAC training parameters are summarized in
Table~\ref{tab:sac_training_configuration}.

\begin{table}
\caption{Parameters used for SAC model training.}
\label{tab:sac_training_configuration}
\setlength{\tabcolsep}{3pt}
\renewcommand{\arraystretch}{1.2}
\begin{tabular}{|p{130pt}|p{100pt}|}
\hline
\textbf{Description} & \textbf{Value} \\
\hline
Actor learning rate &
$5\times10^{-5}$ \\
\hline
Critic learning rate &
$2\times10^{-5}$ \\
\hline
Entropy learning rate &
$1\times10^{-5}$ \\
\hline
Mini-batch size &
$64$ \\
\hline
Replay memory size &
$100000$ \\
\hline
Replay update threshold &
$512$ transitions \\
\hline
Updates per training call &
$32$ \\
\hline
Discount factor $\gamma$ &
$0.99$ \\
\hline
Target update coefficient $\tau$ &
$0.005$ \\
\hline
Initial entropy coefficient $\alpha$ &
$0.2$ \\
\hline
\end{tabular}
\end{table}

\vfill

\end{document}